\documentclass[aps,pre,preprint,superscriptaddress,10pt,twocolumn,showpacs]{revtex4-1}
\usepackage{amssymb}
\usepackage {graphicx}
\usepackage {float}
\usepackage{color}
\usepackage{subfig}
\usepackage{amsmath}
\begin{document}
\title{Ultrashort dark solitons interactions and nonlinear tunneling in the modified nonlinear Schr\"{o}dinger equation with variable coefficients}
 \author{N. M. Musammil}
\altaffiliation{musammilnm7007@gmail.com}
\affiliation{Department of Physics,Calicut University, Malappuram, Kerala, India 673635.}
\author{K. Porsezian}
\altaffiliation{ponzsol@yahoo.com}
\affiliation{Department of Physics, Pondicherry University, Puducherry-605014, India.}
\author{K. Nithyanandan}
\altaffiliation{nithi.physics@gmail.com}
\affiliation{Department of Physics, Pondicherry University, Puducherry-605014, India.}
\affiliation{Laboratoire Interdisciplinaire Carnot de Bourgogne, UMR 6303 CNRS, Univ. Bourgogne Franche-Comt\'e, 9 Av. A. Savary, B.P. 47870, 21078 Dijon Cedex, France.}
\author{P. A. Subha}
\altaffiliation{pasubha@gmail.com}
\affiliation{ Department of Physics, Farook College, Calicut University, Kerala-673632,  India.}
\author{P Tchofo Dinda}
 \affiliation{Laboratoire Interdisciplinaire Carnot de Bourgogne, UMR 6303 CNRS, Univ. Bourgogne Franche-Comt\'e, 9 Av. A. Savary, B.P. 47870, 21078 Dijon Cedex,
 France.}
\begin{abstract}
We present the study of the dark soliton dynamics in an inhomogenous fiber by means of a variable coefficient modified nonlinear Schr\"{o}dinger equation (Vc-MNLSE)  with  distributed dispersion, self-phase modulation, self-steepening and linear gain/loss. The ultrashort dark soliton pulse evolution and interaction is studied by using the Hirota bilinear (HB) method. In particular, we give much insight into the effect of self-steepening (SS) on the dark soliton dynamics. The study reveals a shock wave formation, as a major effect of SS. Numerically, we study the dark soliton propagation in the continuous wave background, and the stability of the soliton solution is tested in the presence of photon noise. The elastic collision behaviors of the dark solitons are discussed by the asymptotic analysis. On the other hand, considering the nonlinear tunneling of dark soliton through barrier/well, we find that the tunneling of the dark  soliton depends on the height of the barrier and the amplitude of the soliton. The intensity of the tunneling soliton either forms a peak or valley and retains its shape after the tunneling. For the case of exponential background, the soliton tends to compress after tunneling through the barrier/well.
\end{abstract}

\maketitle
\section{Introduction}
The  optical soliton is  one of the fascinating technique in the realm of nonlinear fiber optics. In uniform nonlinear fiber, soliton can propagate over a relatively long distance without any considerable attenuation. The exact balancing between the group velocity dispersion (GVD) and the self-phase
modulation (SPM) results in the formation of optical soliton in optical fibers. Soliton was first theoretically predicted by Hasegawa and Tappert
\cite{ref:1} in 1973 and latter experimentally demonstrated by  Mollenauer \emph{et al}. in 1980 \cite{ref:2}. The dynamics of soliton propagation in optical fiber
is governed by the nonlinear Schr\"{o}dinger equation (NLSE). Depending on the signs of GVD, the NLSE admits two distinct types of soliton, namely, bright and dark
solitons. The bright soliton exists in the regime of anomalous dispersion and the dark soliton arises in the regime of normal dispersion. The physics governing the soliton differs depending on whether one considers a bright or a dark soliton, and accordingly features distinct applications \cite{ref:3,ref:4,ref:5,ref:6}.\\

Dark soliton is a localized pulse, which appears as rapid intensity dips on a continuous wave background, unlike the bright counterpart on a zero-intensity background.
Dark soliton has the remarkable stability against the influence of noise and fiber loss.  Based on its inherent stability, it is useful for signal processing,
communications and switching techniques \cite{ref:6}. In many pioneering works, various techniques have been proposed for generating dark soliton in fiber, the
first attempt to the experimental study of dark soliton propagation was made by Emplit \emph{et al.} \cite{ref:7}. The recent advancement of optical technologies
increased the study of dark soliton for various applications, for instance,the dark solitons have been observed in many areas of physics, such as fiber optics
\cite{ref:7,ref:8, ref:9energyyuri,ref:10,ref:11}, plasma \cite{ref:12,ref:13}, waveguide arrays \cite{ref:14}, Bose-Einstein condensates \cite{ref:15,ref:16},
water surface \cite{ref:17}, etc.\\

In several experiments on optical soliton propagation in fibers, the output pulse has been found to be asymmetric due to the self-steepening (SS) effect.  SS is
found to be crucial in optical communication system, especially in the ultrashort pulse propagation in long distance optical fibers system.  The modified nonlinear Schr\"{o}dinger equation (MNLSE) describing the soliton propagation with SS effect \cite{ref:18,ref:19,ref:20,ref:21,ref:22sscw} has been under considerable interest over a long time. To study the MNLSE, many mathematical techniques have been demonstrated and large class of analytical solutions
were discussed in \cite{ref:23,ref:24,ref:25}. All those investigations focused on the MNLSE model with constant coefficient, considering an ideal optical fiber
transmission system. However, as a result of non-uniformities, influenced by the spatial variations of the fiber parameters, the realistic optical fiber medium
exhibits inhomogeneous behavior. The variable coefficient NLS model may serve as a practical model for describing the soliton dynamics in inhomogeneous systems
\cite{ref:26,ref:27,ref:28,ref:29,ref:30,ref:31}. In this work,  we consider the following variable coefficient MNLS equation (Vc-MNLSE), which governs the
ultrashort dark soliton propagation in an inhomogeneous fiber with the distributed dispersion, nonlinearity, SS and linear gain/loss \cite{ref:32,ref:33workppr,ref:34}:\\

\begin{equation}\label{vc-mnlse}
    iq_\xi-\frac{1}{2}D(\xi)q_{\tau\tau}+R(\xi)(q|q|^2)+iS(\xi)(q|q|^2)_\tau+ip(\xi)q=0
\end{equation}
\noindent
where, $q(\xi,\tau)$ is the complex amplitude of the pulse envelope, the variables $\xi$ and $\tau$ represent the normalized spatial and temporal coordinates.
The GVD, Kerr nonlinearity, SS and amplification/absorption effects are related to the coefficients $D(\xi)$, $R(\xi)$, $S(\xi)$ and $p(\xi)$, respectively.\\

Many mathematical techniques have been proposed to study the characteristics of the pulse evolution in nonlinear optical fibers, including perturbation
theory \cite{ref:35perturbation}, numerical simulation \cite{ref:36numerics}, variational approach \cite{ref:37variational} etc. In this paper, we use Hirota's
bilinear method, which is one of the famous analytical method to investigate the multi-soliton propagation in fibers.  Hirota's bilinear method was a perturbation
technique to get the soliton solution and provides explicit analytical expressions for the soliton pulses. The important physical quantities and interaction
behaviors can be well understood with the use of this method \cite{ref:38,ref:39,ref:40,ref:41}.\\

The bright-soliton propagation and the variation of bright soliton energy due to self phase modulation (SPM) and SS in Vc-MNLSE has been studied in \cite{ref:32}. The dark and anti-dark
solitons propagation have been discussed in \cite{ref:33workppr}.  In this paper, we report a more general form of dark soliton solutions for Vc-MNLSE with the
conventional form of bilinear transformation as in Refs.~ \cite{ref:42,ref:43,ref:44,ref:45,ref:46,ref:47}. Such study has not been discussed in the context of
Vc-MNLSE.  By using this approach, we  exclusively studied the ultrashort dark soliton dynamics with various  inhomogeneous effects, such as pulse amplification/absorption,
compression, boomerang soliton, dispersion-managed transmission systems and nonlinear tunneling. In addition to that, we have studied the impact of SPM and SS
effect on the pulse energy, and by using asymptotic analysis we observed the energy conservation of dark soliton during an elastic collision. Moreover, by the
direct numerical simulation, we studied the shock formation of pulse under the influence of SS effect.\\

In this work,  we have also paved much attention on nonlinear (NL) tunneling effect.  Out of other effects, NL tunneling has been under considerbale interest especially
in the context of optical switching.  Recently, many experimental and theoretical works was devoted to study the NL tunneling effect of solitons in different
physical systems \cite{ref:48,ref:49,ref:50,ref:tunexp}. The  NL tunneling of bright and dark soliton in the various forms of  NLS model has been investigated.
In this paper, we study the NL tunneling of ultrashort dark soliton for the first time to be best of our knowledge.\\

The remaining of the paper is organized as follows.  Sec.2 presents the exact dark soliton solutions by the Hirota's bilinear method. In Sec. 3, the one soliton solution and the influence of SS in the soliton dynamics has been presented.  Sec.4 presents the two solitons solution and asymptotic analysis to study the collision behavior. A brief discussion about the various physical effects involved in the dynamics of dark soliton propagation through inhomogenous fiber is presented in Sec. 5. The tunneling
of dark soliton through barrier/well is discussed in Sec.6, followed by a brief summary and conclusion in Sec. 7.
\section{Exact dark soliton solutions by Hirota method}
In this section, we  use Hirota's bilinear method to investigate the analytical dark soliton solutions of Eq.  (\ref{vc-mnlse}). Here, we use the transformation
as followed in Refs.~\cite{ref:42,ref:43,ref:44,ref:45,ref:46,ref:47}, which is expected to give an exact form of dark soliton solutions. By using this
transformation, the nonlinear differential equations can be transformed into bilinear differential equations. Then, with the different levels of perturbation
expansion, the exact form of dark soliton solutions can be derived.\\

In order to construct the dark soliton solutions, we apply the following form of Hirota bilinear transformation;
   \begin{align}\label{3}
   q(\xi,\tau)=g(\xi)\frac{G}{F}
   \end{align}
 where, G is a complex function and F is a real function. By substituting this transformation into Eq. (\ref{vc-mnlse}), the following bilinear equations can be
 obtained,
\small
\begin{align}\label{4}
[iD_\xi-\frac{1}{2}D(\xi)D_\tau^2+\lambda(\xi)](G.F)&=0
\end{align}
\begin{align}\label{5}
\delta|G|^2+ D_\tau^2(F.F)&+i\gamma (D_\tau(G^*.G)+3\frac{G^*}{F}D_\tau(G.F))= \frac{2\lambda(\xi)}{D(\xi)}F^2
\end{align}
\normalsize
with the condition  $g_z(z)+g(z)p(z)=0$. Here, $\lambda(z)$ is an analytic function to be determined,  $\delta$ and $\gamma$ can be introduced as, $\delta = \frac {2R(\xi)}{D(\xi)}g(\xi)^2$, $\gamma = \frac {2S(\xi)}{D(\xi)}g(\xi)^2 $.  $D_z$ and $D_t$ are the bilinear differential operators \cite{ref:29} defined by
\small
\begin{multline*}
D_z^m D_t^n(g.f)=(\frac {\partial}{\partial z}-\frac {\partial}{\partial z'})^m (\frac {\partial}{\partial t}-\frac {\partial}{\partial t'})^n
g(z,t)f(z,t)|_{z'=z, t'=t}
\end{multline*}
\normalsize
By solving the above set of equations (\ref{4})-(\ref{5}), we consider the power series expansion of G and F as,
\begin{align}
 G &= g_0[1+\sum_{n=1}^\infty \varepsilon^n g_n(\xi,\tau)] \\
 F &= 1+\sum_{n=1}^\infty \varepsilon^n f_n(\xi,\tau)
\end{align}
with $\varepsilon$ as the formal expansion parameter. While applying Hirota Direct method, we assume $g_0$, $g_n$ and $f$ as polynomials of exponential
functions.
\section{One-soliton solutions}
In order to get the dark one-soliton solution, the power series expansions for $G$ and $F$ are truncated corresponding to the lowest order in $\epsilon$ as
follows, $G=g_0(1+g_1)$ and  $F=1+f_1$. Then, back to bilinear equations  (\ref{4})-(\ref{5}), we obtain

\begin{figure}[h!]
\begin{center}
\includegraphics[height=4 cm, width=7cm]{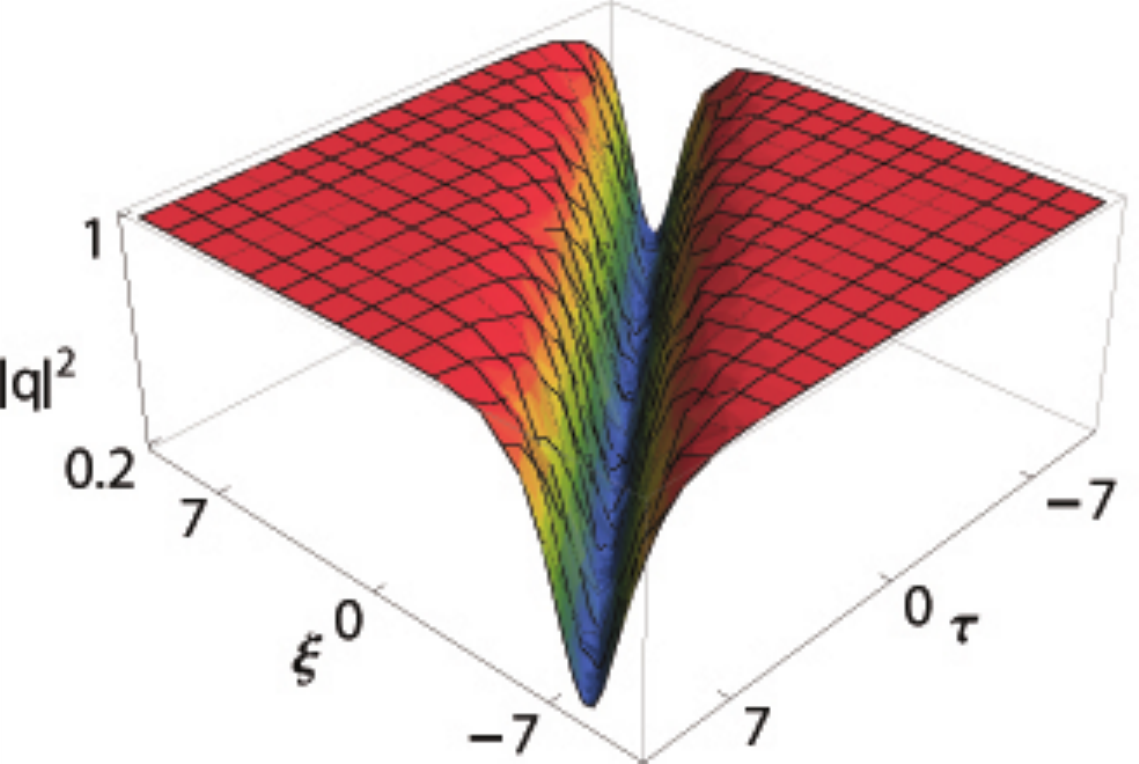}
\caption{(Color online) The dark one-soliton propagation through homogenous fiber for parameters $k_1=1.5$, $k = a_0 = D(\xi) = 1$, $p=0$, $\delta= 2$ and
$\gamma=1$}
\label{dark one soliton}
\end{center}
\end{figure}

\begin{widetext}
 \begin{align*}
  g_0 & = a_0 e^{i\psi}, & g_1 &=\alpha_1  e^{\theta_1},  &
   f_1 &= e^{\theta_1} \\ g(\xi) &= e^{-\int p(\xi)d\xi} ,&
      \psi &=k \tau-\omega\int D(\xi)d\xi ,& \theta_1& =k_1\tau-\omega_1\int D(\xi)d\xi \\
       \lambda &=  \frac{ a_0^2}{2}[\delta-\gamma k]D(\xi) ,&  \omega  &=-  \frac{\lambda}{D(\xi)}-\frac{ k^2}{2},&
       \alpha_1 &=\frac{2\omega_1+2 kk_1 +i k_1^2}{2\omega_1+2 kk_1-i k_1^2}
  \end{align*}
  \begin{multline*}
       \omega_1 = \frac{1}{12 k \gamma  a_0^2-4 k_1^2} (4k k_1^3+\gamma  a_0^2 k_1(-12k^2-3ikk_1+k_1^2)\\-\sqrt{(-k_1^2(4k_1^4+4a_0^2
       k_1^2(k\gamma-2\delta+3i\gamma k_1)-\gamma  a_0^2(39 k^2 \gamma-24k\delta+30ik\gamma k_1+\gamma k_1^2 ) ))})
\end{multline*}
\end{widetext}
\normalsize
The one-soliton solution can be written as,
\begin{equation}\label{one}
 q(\xi,\tau)= \frac{a_0[ (1+\alpha_1)+(\alpha_1-1)tanh(\frac{\theta_1}{2})]}{2e^{\int p(\xi)d\xi}  e^{-i\psi}}
 \end{equation}
Here, $a_0 e^{i\psi}$ represents the background wave solution. $a_0$ and $\psi$ are real functions denoting the amplitude and phase of the background wave.
Using Eq.  (\ref{one}), the propagation of  dark one-soliton through homogenous fiber is depicted in the Fig. \ref{dark one soliton}. From the Eq. (\ref{one}),
we can analyze the dynamics of dark soliton pulse in inhomogeneous fibers.\\

In order study the dynamics of dark soliton and to characterize the inhomogeneous features of propagating optical dark soliton, some of the physical quantities
such as velocity, width, amplitude and energy are important. Such quantities can be defined as follows,
$$ V=\frac{\omega_1}{\kappa_1}D(\xi), \quad W=\frac{ 1}{\kappa_1}, \quad A=  |\frac{a_0(1+\alpha_1 )}{2 e^{\int p(\xi)d\xi}}|$$\\
The energy E and power P, in terms of the background amplitude $a_0$ can be expressed as $E= \int_{-\infty}^\infty P d\tau$ and $P(\xi,\tau)=a_0^2-|q|^2$,
respectively. Here, the instantaneous power is obtained as a difference between the total power and the corresponding value for the background \cite{ref:60}. The
energy, corresponding to the one-soliton solution as given by the Eq. (\ref{one}), can be written as
 \begin{equation}\label{oneenergy}
 \quad E= \int_{-\infty}^\infty (a_0^2-\mid q\mid^2)d\tau= \frac{2-  \alpha_1-\alpha_1^*}{  a_0^{-2} e^{2\int p(\xi)d\xi} k_1}
 \end{equation}
From the above set of equations, we can analyze the effect of inhomogeneity on the physical quantities of dark soliton. It is interesting to note that, $p(\xi)$
affects the soliton amplitude and energy, $D(\xi)$ affects the soliton velocity. The soliton width is related to the wave number $k_1$.

\begin{figure*}[htb!]
\subfloat[]{\label{sswt1}\includegraphics[width = 7cm]{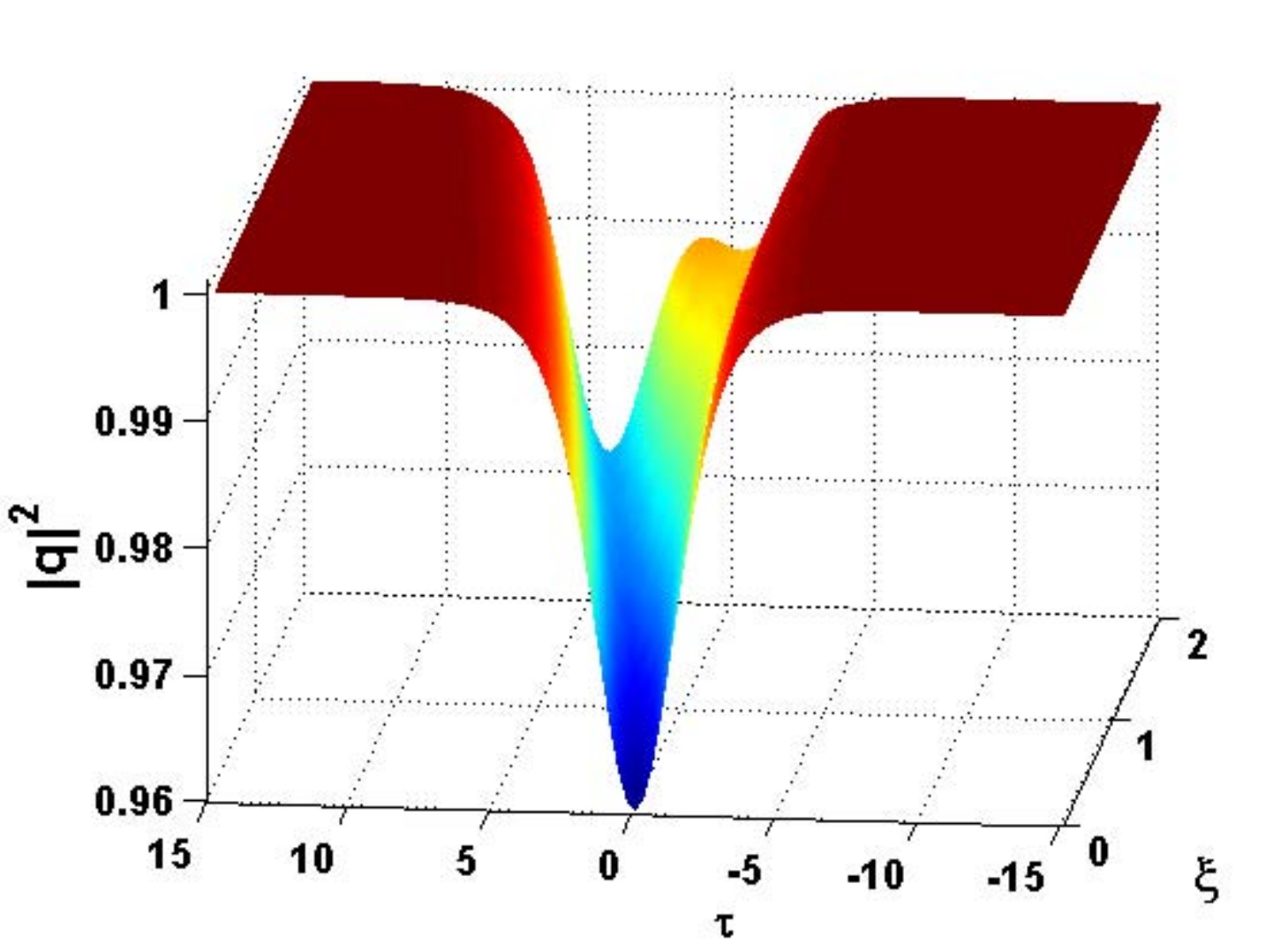}}
\subfloat[]{\label{sswt2}\includegraphics[width = 7cm]{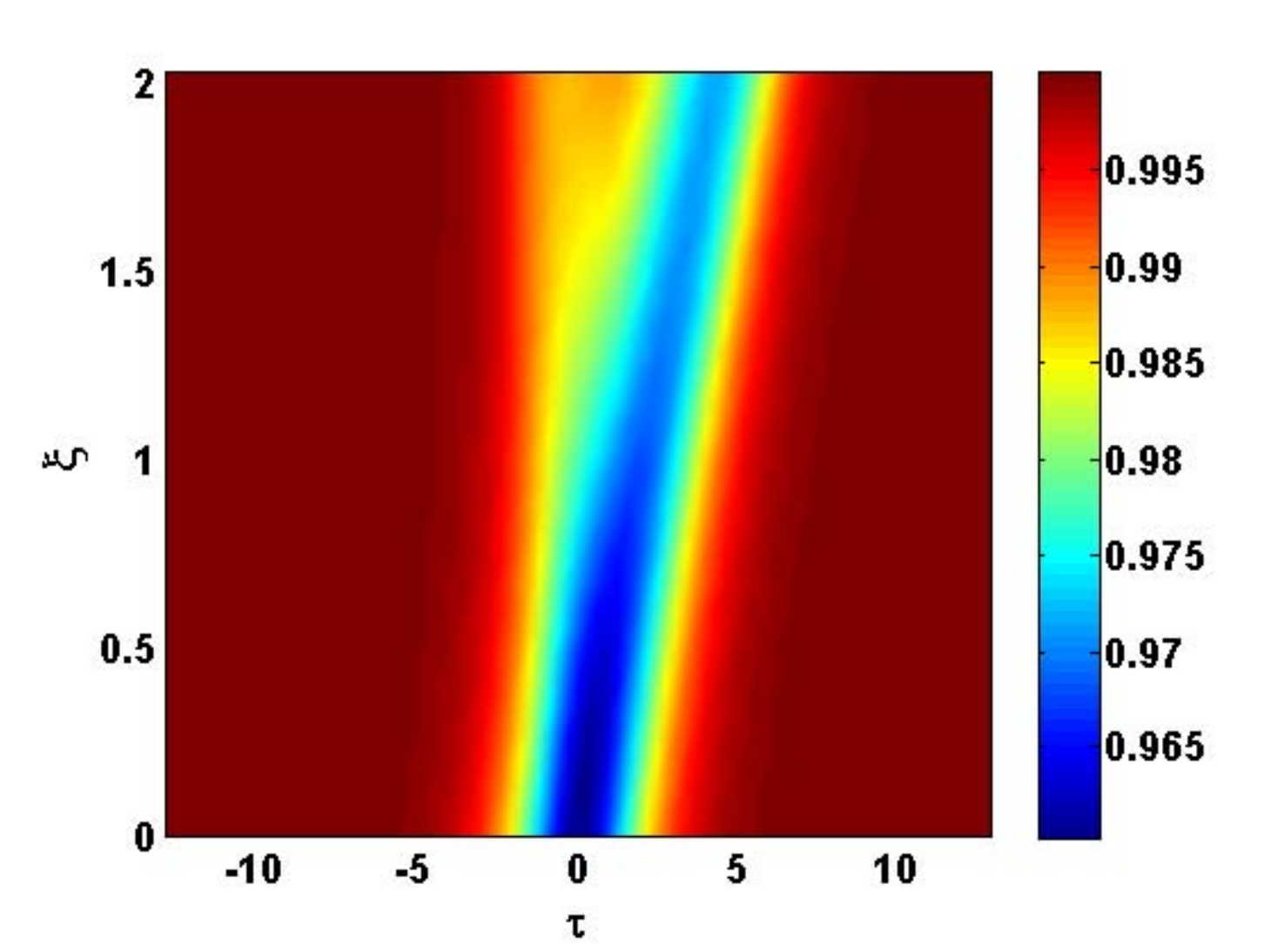}}\\
\subfloat[]{\label{ssw1}\includegraphics[width = 7cm]{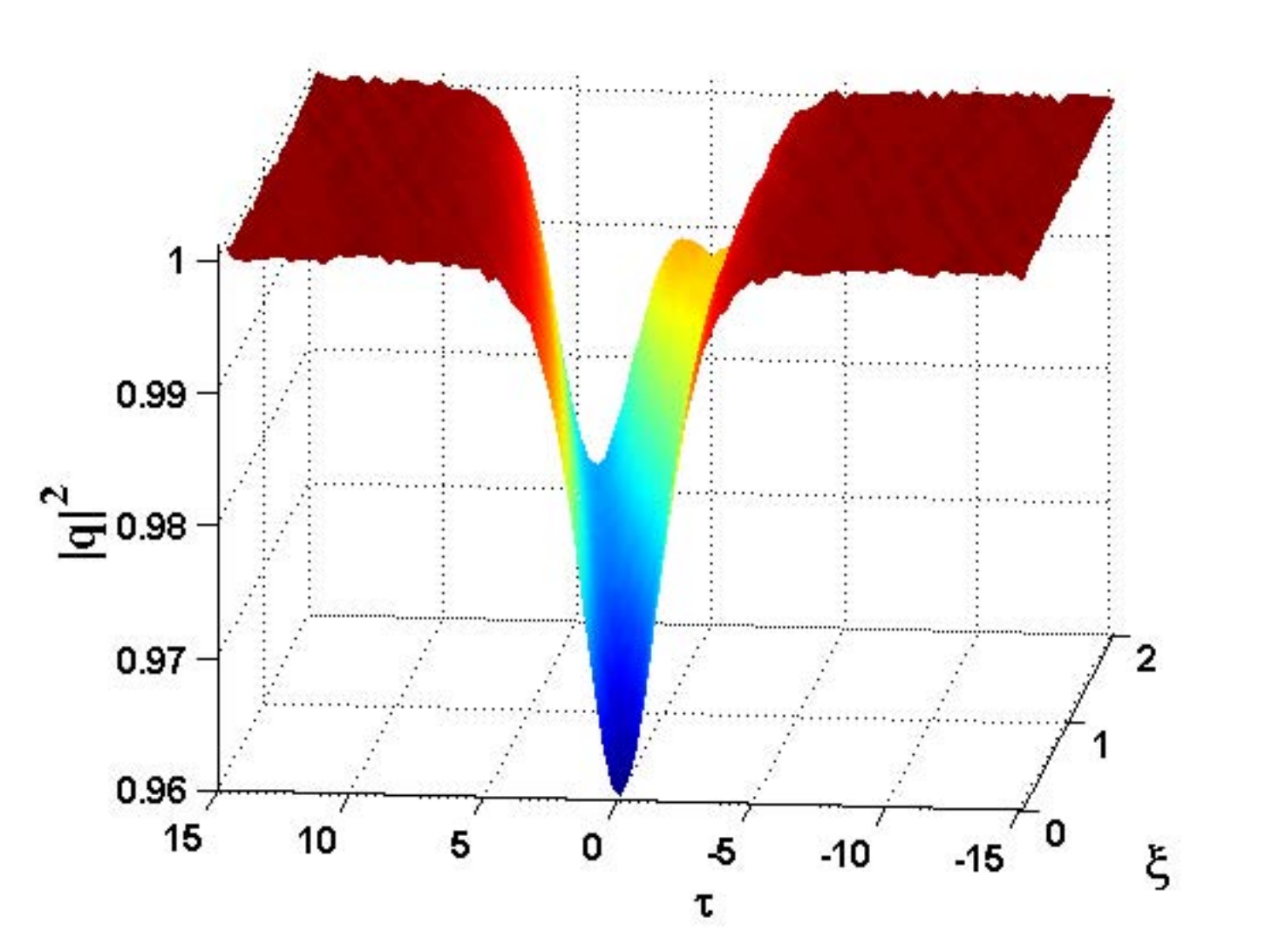}}
\subfloat[]{\label{ssw2}\includegraphics[width = 7cm]{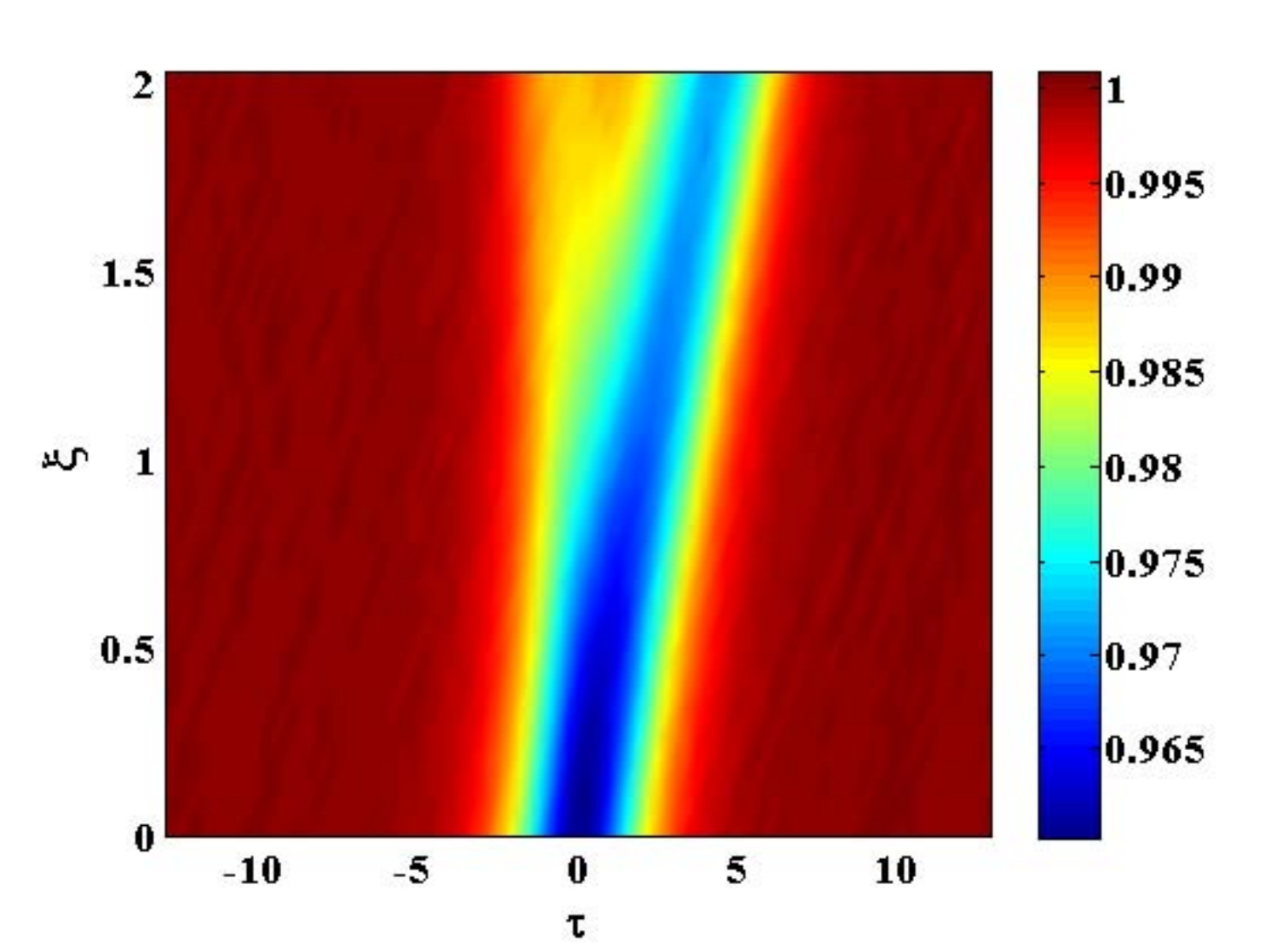}}
\caption{(Color online) Figs (a) and (b) show the dark soliton evolution and the formation of shcok wave. Figs (c) and (d) shows the stable propagation of the soliton and the shock wave formation in the presence of strong photon noise. The parameters of relevant physical quantities are $k_1=k = a_0 =1$ $\delta= 2$, $\gamma=1$and $p(\xi)= 0$.}
\label{sseffect}
\end{figure*}

\subsection{Direct numerical simulation}

One of the essential aspect of a solitary wave is its stability on propagation. Unlike the conventional pulses of different form, the solitons are
relatively stable, even in an environment subjected to external perturbations. Hence, in order to validate the signature of soliton, such as stable
propagation over appreciable distance, and the stability against perturbation, we perform numerical simulation using split-step Fourier method. In order
to check the solution stability of our dark soliton solutions, as a representative case, we consider the one soliton solution given by  Eq. (\ref{one}), and perform
the stability analysis in two parts, (i) direct numerical simulation of propagation of soliton using Vc-MNLSE, and (ii) the propagation of soliton subject to
perturbation such as the photon noise.   Fig. \ref{sseffect} shows the numerical simulation of stable propagation of the dark soliton in the continuous back ground.  Fig. \ref{sswt1} shows the stable propagation of soliton pulse, and followed the shock wave formation as a consequence of SS due to the intensity dependent group velocity \cite{ref:sschina}. As of now, the propagation of soliton pulses have been considered in an ideal environment. However, there are numerous effects can contribute to instability in the soliton propagation. Therefore, it is very informative to study the stability of the soliton in an environment subject to external noise or perturbations. To this end, we generated a photon noise, which corresponds to $0.35\%$ of the continuous background. This is indeed an appreciable noise level, which can potentially perturb any propagation, as evident from the smooth pulse shown in Fig. \ref{sswt1}  and the noisy pulse depicted in Fig. \ref{ssw1}.  So, the initial condition for the simulation is the soliton profile with strong perturbation.  Fig. \ref{ssw1} shows the simulation results for the same parameters as chosen earlier. It is very evident that the dark soliton show remarkable stability against strong perturbation. The formation of the shock wave can also be clearly observed in the simulation. Thus, one can draw out a conclusion that the dark soliton solution constructed through the Hirota method shows excellent stability, which has been
confirmed through direct numerical simulations.


\section{ Two-soliton solutions}
\begin{figure}[h!]
\begin{center}
\includegraphics[height=4 cm, width=7cm]{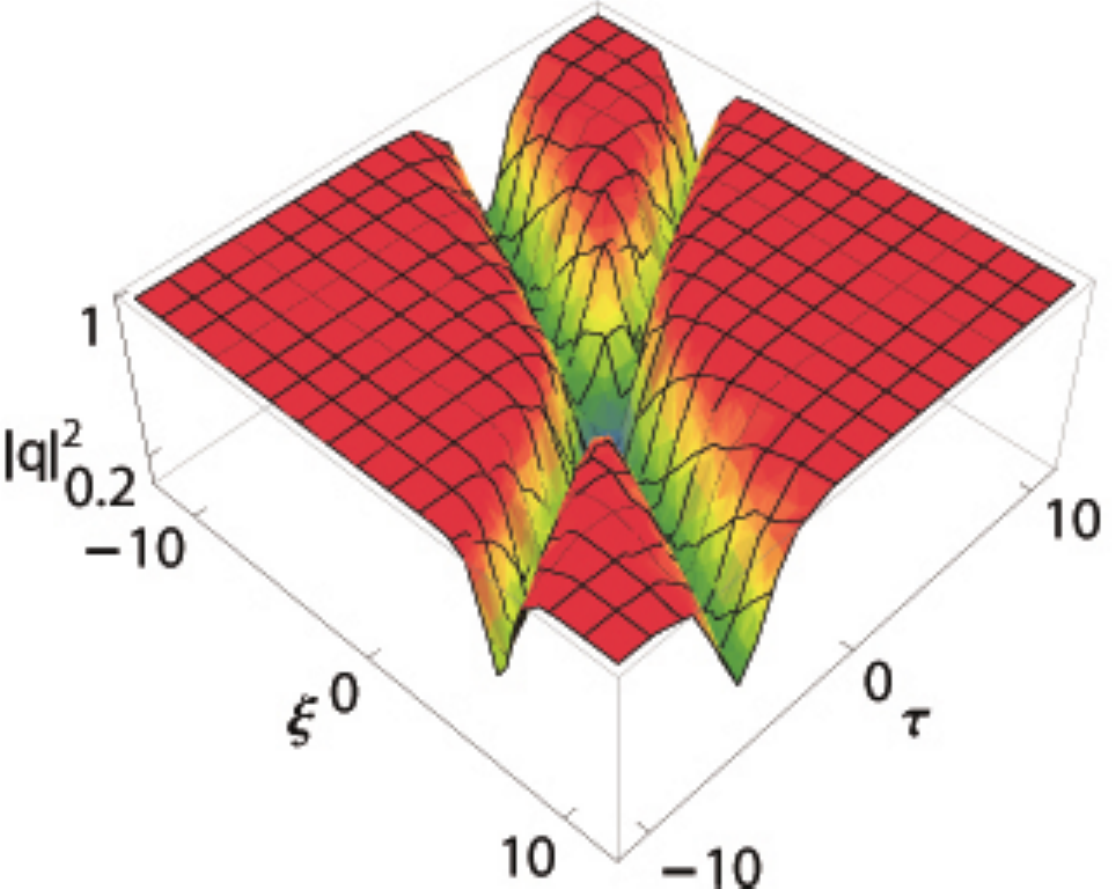}
\caption{(Color online) The dark two-soliton propagation through homogenous fiber for parameters $k_1=-1.5$, $k_2=1.5$, $k = a_0 = D(\xi) = 1$, $p=0$,
$\delta=1$ and $\gamma=0.2$.}
\label{dark two soliton}
\end{center}
\end{figure}

To get the dark two-soliton solution, the power series expansions for $G$ and $F$ are truncated as follows, $G=G_0(1+g_1+g_2)$ and  $F=1+f_1+f_2$. Then, back to
bilinear Eqs.  (\ref{4})-(\ref{5}), we obtain
\begin{widetext}
\begin{align*}
g_0 &=a_0 e^{ik\tau-i\omega\int D(\xi)d\xi} , & g(\xi)&= e^{-\int p(\xi)d\xi},&
\lambda &=   \frac{ a_0^2}{2}[\delta-\gamma k]D(\xi) ,&   \omega  &=-  \frac{\lambda}{D(\xi)}-\frac{ k^2}{2}
\end{align*}
\begin{align*}
g_1&=\alpha_1\, e^{\theta_1}+\alpha_2 e^{\theta_2 }, & f_1&= e^{\theta_1}+e^{\theta_2},&
g_2&=A_{12}\alpha_1\alpha_2\, e^{\theta_1 +\theta_2 }, & f_2&= A_{12}e^{\theta_1 +\theta_2}
\end{align*}
\begin{align*}
 \theta_1 &=k_1 \tau-\omega_1\int D(\xi)d\xi , &
 \theta_2& =k_2\tau-\omega_2\int D(\xi)d\xi , &
 \alpha_1 &=\frac{2\omega_1+2 kk_1 +i k_1^2}{2\omega_1+2 kk_1-i k_1^2} , &
  \alpha_2 &=\frac{2\omega_2+2 kk_2 +i k_2^2}{2\omega_2+2 kk_2-i k_2^2}\\
  \end{align*}
  \begin{multline*}
 \omega_1 = \frac{1}{12 k \gamma  a_0^2-4 k_1^2} (4k k_1^3+\gamma  a_0^2 k_1(-12k^2-3ikk_1+k_1^2)\\ -\sqrt{(-k_1^2(4k_1^4+4a_0^2 k_1^2(k\gamma-2\delta+3i\gamma
 k_1)-\gamma  a_0^2(39 k^2 \gamma-24k\delta+30ik\gamma k_1+\gamma k_1^2 ) ))})
 \end{multline*}
 \begin{multline*}
 \omega_2 =\frac{1}{12 k \gamma  a_0^2-4 k_1^2} (4k k_1^3+\gamma  a_0^2 k_1(-12k^2-3ikk_1+k_1^2)\\ -\sqrt{(-k_1^2(4k_1^4+4a_0^2 k_1^2(k\gamma-2\delta+3i\gamma
 k_1)-\gamma  a_0^2(39 k^2 \gamma-24k\delta+30ik\gamma k_1+\gamma k_1^2 ) ))})
\end{multline*}
\begin{align*}
 A_{12}=-\frac{2i(\alpha_1-\alpha_2)(\omega_2-\omega_1-k k_1+k k_2)-(\alpha_1+\alpha_2)(k_1-k_2)^2}{2i(1-\alpha_1\alpha_2)(\omega_1+\omega_2+k k_1+k
 k_2)-(\alpha_1\alpha_2+1)(k_1+k_2)^2}
 \end{align*}
 \end{widetext}
 The two-soliton solution can be written as,
 \begin{equation}\label{two}
 q(\xi,\tau)=e^{-\int p(\xi)d\xi}\frac{g_0(1+g_1+g_2)}{(1+f_1+f_2)}
 \end{equation}
 Using Eq. (\ref{two}), the propagation of dark two-soliton through homogenous fiber is depicted in the Fig. (\ref{dark two soliton})

\subsection{Two-soliton interactions}
The interaction behaviors between two solitons in fibers can be revealed by the asymptotic states of soliton solution. Based on the two-soliton solution,
we discuss the collision between dark solitons in inhomogeneous fibers.  The asymptotic analysis of two soliton solutions are constructed as follows:\\

 \small
 1) Before collision\\\\
(a)$ S_1^-(\theta_1\sim 0, \theta_2\rightarrow -\infty)$
\begin{equation}\label{bc1}
q(\xi,\tau)\rightarrow S_1^-= \frac{a_0 e^{i\psi}}{2 e^{\int p(\xi)d\xi}} [(1+\alpha_1)+(\alpha_1-1) tanh(\frac{\theta_1}{2})]
\end{equation}
(b)$ S_2^-(\theta_2\sim 0, \theta_1\rightarrow \infty)$
\begin{equation}\label{bc2}
q(\xi,\tau)\rightarrow S_2^-=  \frac{a_0 \alpha_1 e^{i\psi}}{2 e^{\int  p(\xi)d\xi}} [(1+\alpha_2)+(\alpha_2-1) tanh(\frac{\theta_2}{2}+\ln(\sqrt{A_{12}}))]
\end{equation}
2)After collision\\\\
(a)$ S_1^+(\theta_1\sim 0, \theta_2\rightarrow \infty)$
\begin{equation}\label{ac1}
q(\xi,\tau)\rightarrow S_1^+=   \frac{a_0 \alpha_2 e^{i\psi}}{2 e^{\int  p(\xi)d\xi}} [(1+\alpha_1)+(\alpha_1-1) tanh(\frac{\theta_1}{2}+\ln(\sqrt{A_{12}}))]
\end{equation}
(b)$ S_2^+(\theta_2\sim 0, \theta_1\rightarrow -\infty)$
\begin{equation}\label{ac2}
q(\xi,\tau)\rightarrow S_2^+= \frac{a_0 e^{i\psi}}{2 e^{\int  p(\xi)d\xi}} [(1+\alpha_2)+(\alpha_2-1) tanh(\frac{\theta_2}{2})]
\end{equation}
\normalsize
From the asymptotic expressions before collision (\ref{bc1})- (\ref{bc2}) and after collision  (\ref{ac1})- (\ref{ac2}), one can infer the elastic interaction
and particle like behavior of solitons during the time of collisions between $S_1$ and $S_2$. The relevant physical quantities of solitons $S_1$ and $S_2$ before
and after collisions are mentioned in Table 1.

\begin{table*}[htb]
\caption{Physical quantities of solitons $S_1$ and $S_2$ before and after the collision.} 
\centering 
\begin{tabular}{c c c c c c} 
\hline\hline 
 Solitons & Velocities & Widths & Amplitudes & Energies \\ [0.5ex] 
\hline 
$s_1^- $& $\frac{\omega_1}{k_1}D(\xi) $& $\frac{ 1}{k_1}$  & $   |\frac{a_0(1+\alpha_1 )}{2 e^{\int p(\xi)d\xi}}|$& $ \frac{a_0^2( 2- \alpha_1-\alpha_1^*)}{ k_1
e^{2\int p(\xi)d\xi}}$    \\ \\
$s_2^- $& $\frac{\omega_2}{k_2}D(\xi) $ &   $\frac{1}{k_2}$ &  $ | \frac{a_0 \alpha_1( 1+ A_{12}\alpha_2 )}{ e^{\int p(\xi)d\xi}(1+A_{12})} |$& $ \frac{ a_0^2(2-
\alpha_2-\alpha_2^*)}{ k_2 e^{2\int p(\xi)d\xi}}  $ \\ \\
$s_1^+$ & $\frac{\omega_1}{k_1}D(\xi) $ &  $\frac{ 1}{k_1}$ &  $  | \frac{a_0 \alpha_2( 1+ A_{12}\alpha_1 )}{ e^{\int p(\xi)d\xi}(1+A_{12})}  |$& $  \frac{
a_0^2(2- \alpha_1-\alpha_1^*)}{ k_1 e^{2\int p(\xi)d\xi}}  $   \\\\
$s_2^+ $ & $\frac{\omega_2}{k_2}D(\xi) $ &  $\frac{ 1}{k_2}$ &  $  | \frac{a_0(1+\alpha_2 )}{2 e^{\int p(\xi)d\xi}}|$ & $ \frac{ a_0^2 (2-\alpha_2-\alpha_2^*)}{
k_2 e^{2\int p(\xi)d\xi}}$ \\
  [2ex] 
\hline 
\end{tabular}
\label{table:nonlin}
\end{table*}

\begin{figure}[ht!]
\subfloat[]{\includegraphics[width = 4cm]{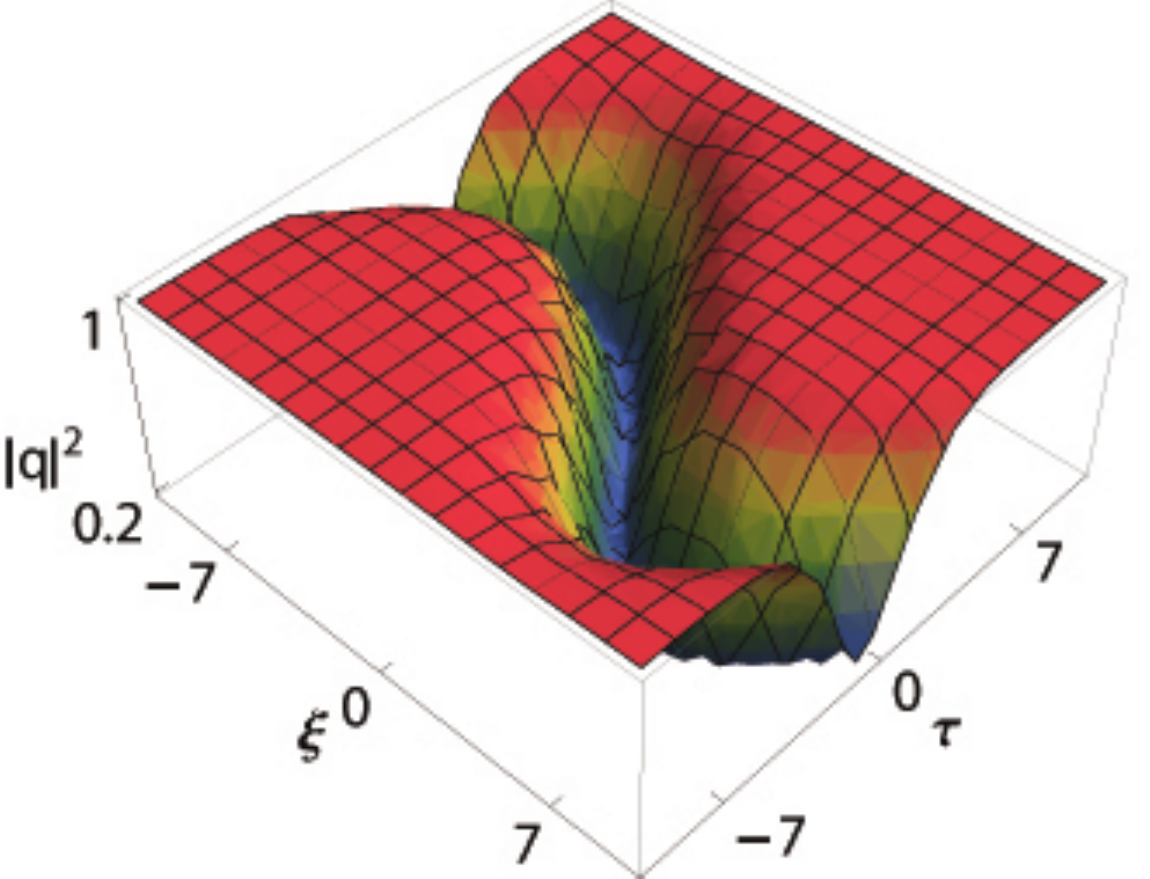}}
\subfloat[]{\includegraphics[width = 4cm]{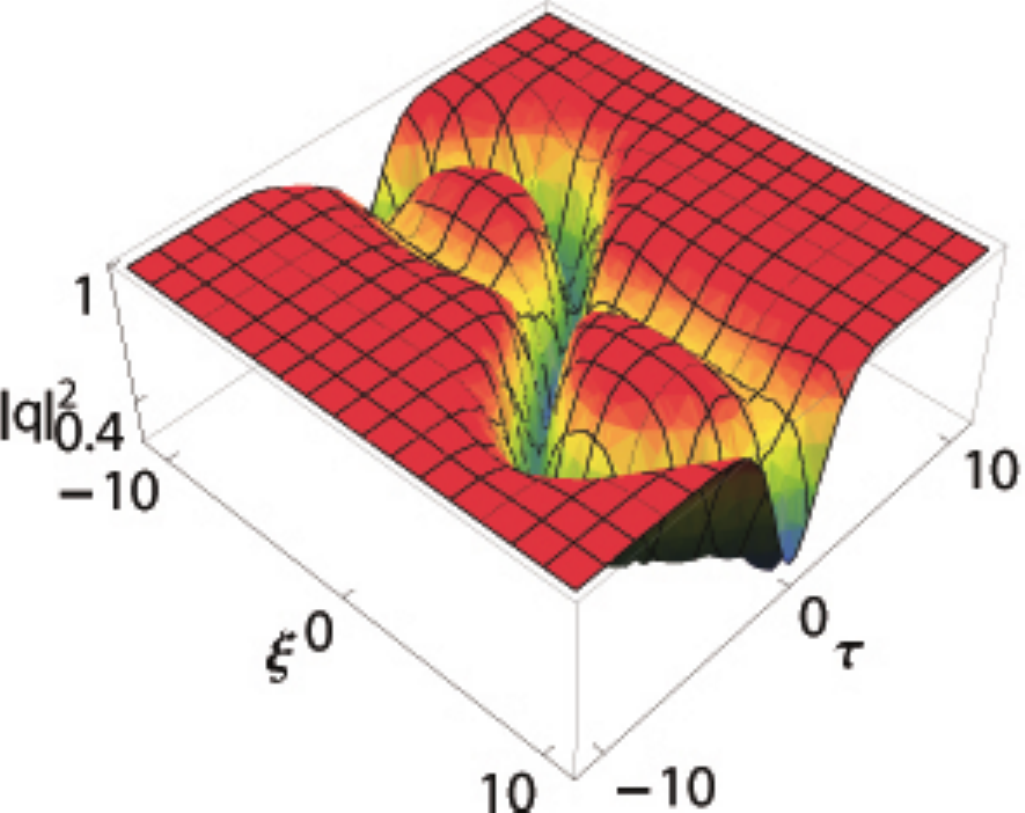}}
\caption{(Color online) The dispersion and nonlinearity managed dark solitons, (a) One-soliton (b) Two-soliton. Other physical quantities are $k = a_0 = D(\xi) =
Cos(0.3\xi)$, $p=0$, $\delta= 2$ and $\gamma=1$.}
\label{periodic}
\end{figure}

\begin{figure} [ht!]
\subfloat[]{\includegraphics[width = 4cm]{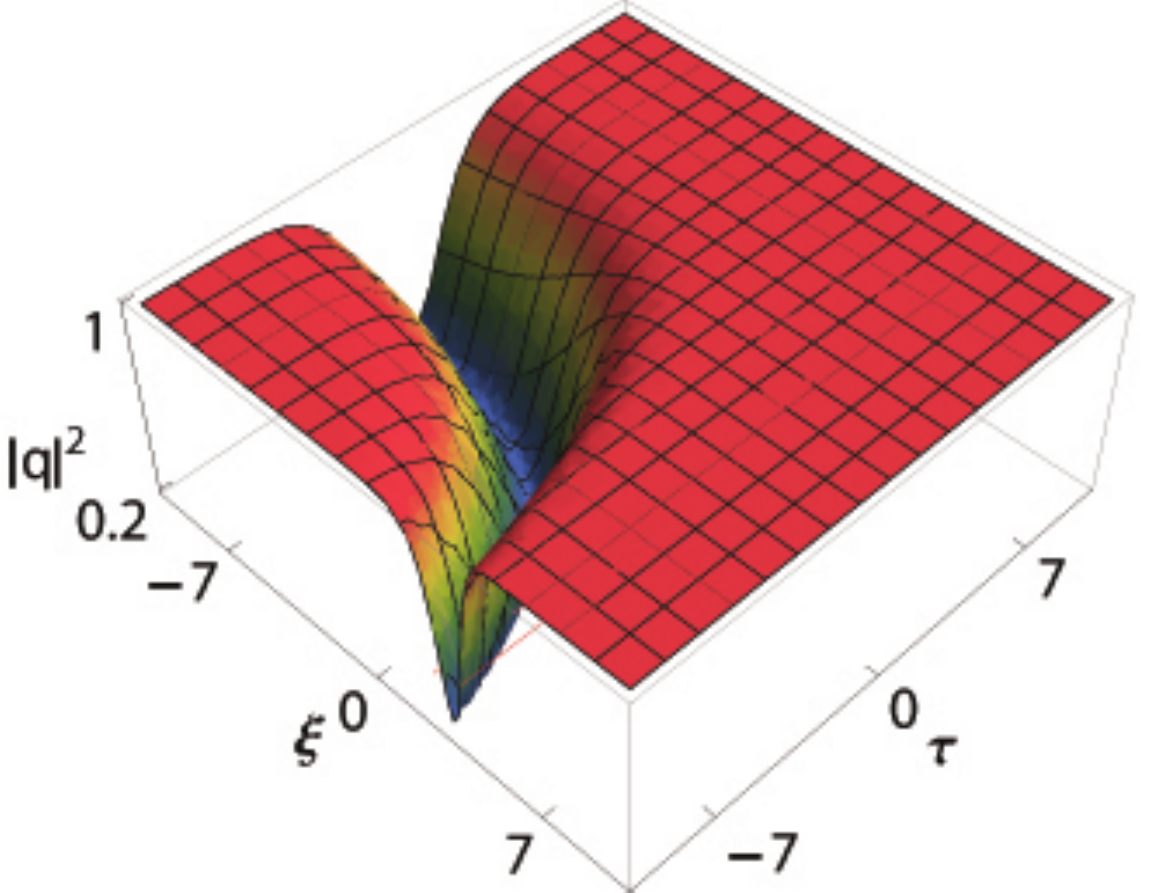}}
\subfloat[]{\includegraphics[width = 4cm]{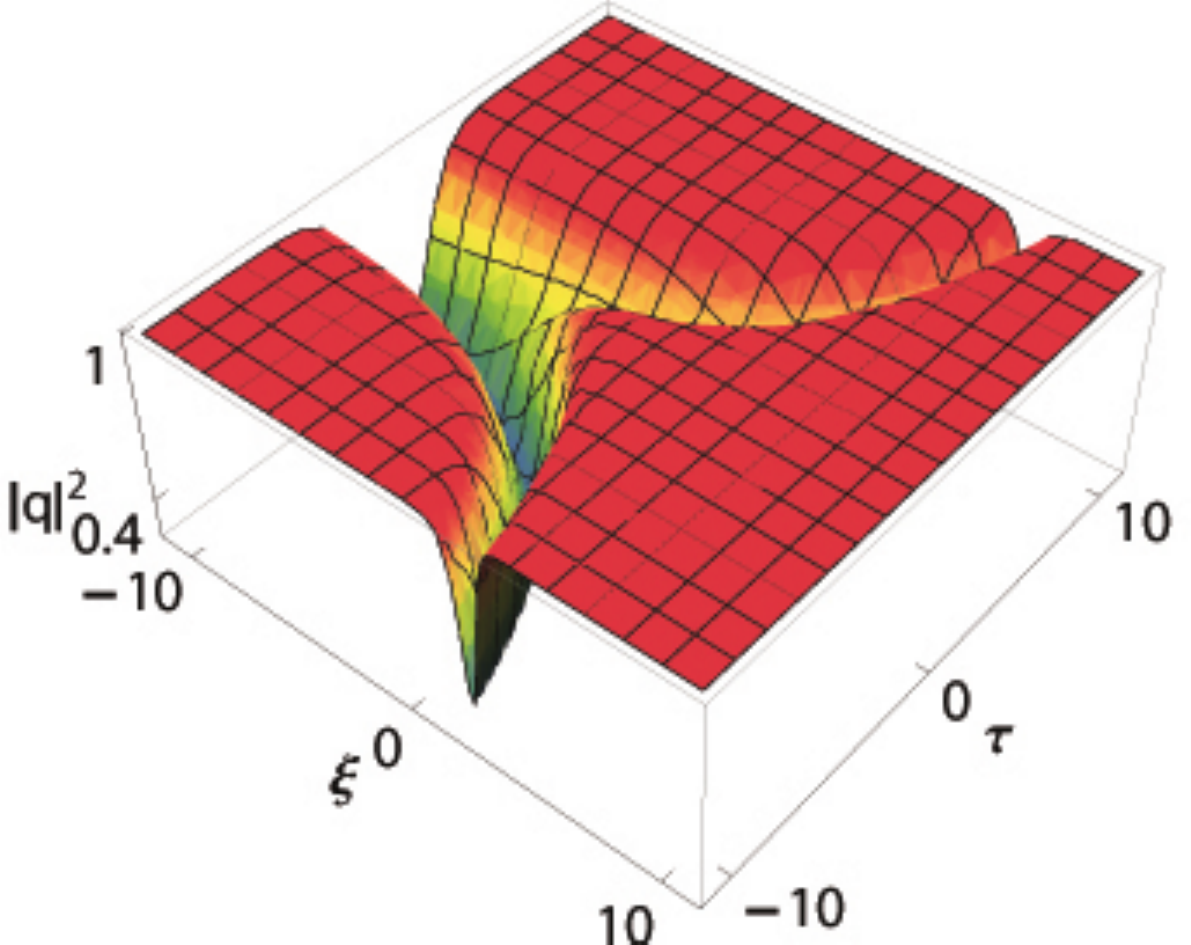}}
\caption{(Color online) The pulse compression of dark solitons, (a) One-soliton (b) Two-soliton. Other physical quantities are $k = a_0 = D(\xi) = Exp(0.3\xi)$,
$p=0$, $\delta= 2$ and $\gamma=1$.}
\label{Compression}
\end{figure}
\begin{figure}[ht!]
\subfloat[]{\includegraphics[width = 4cm]{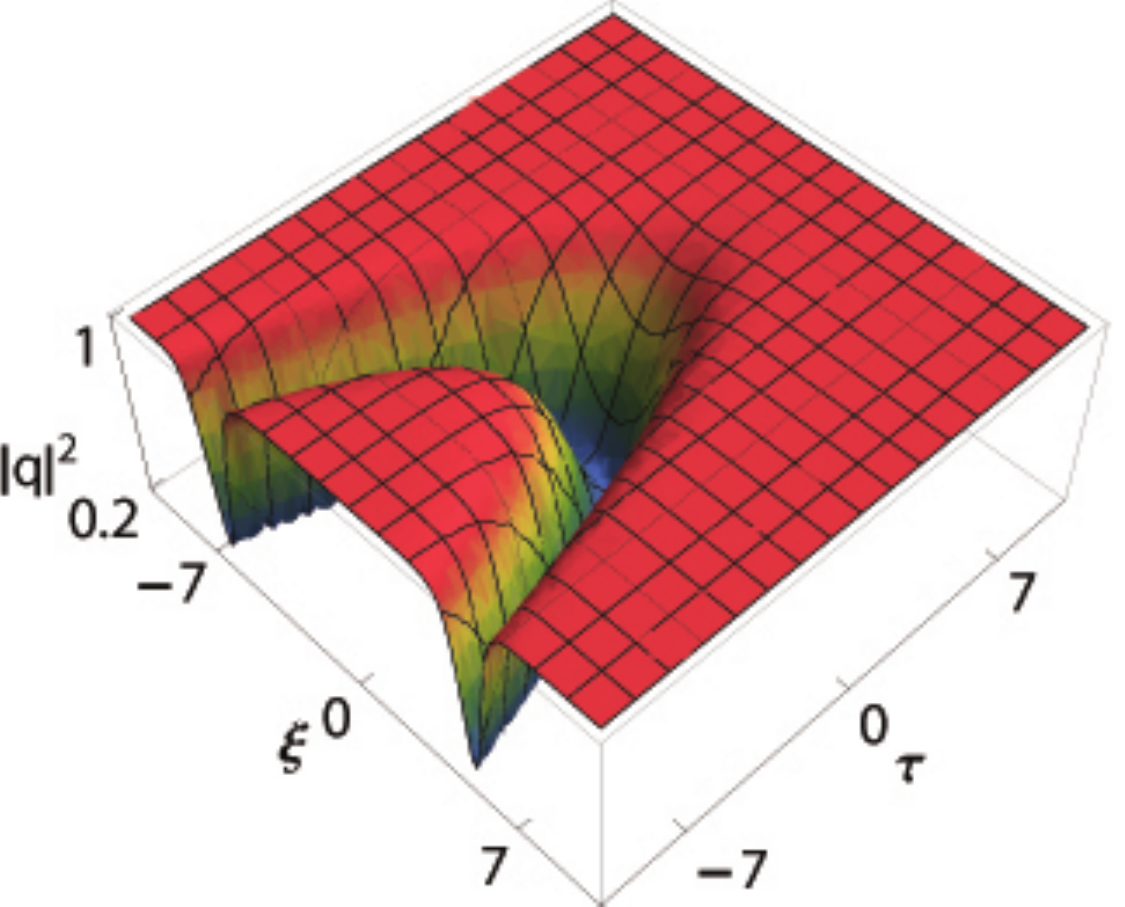}}
\subfloat[]{\includegraphics[width = 4cm]{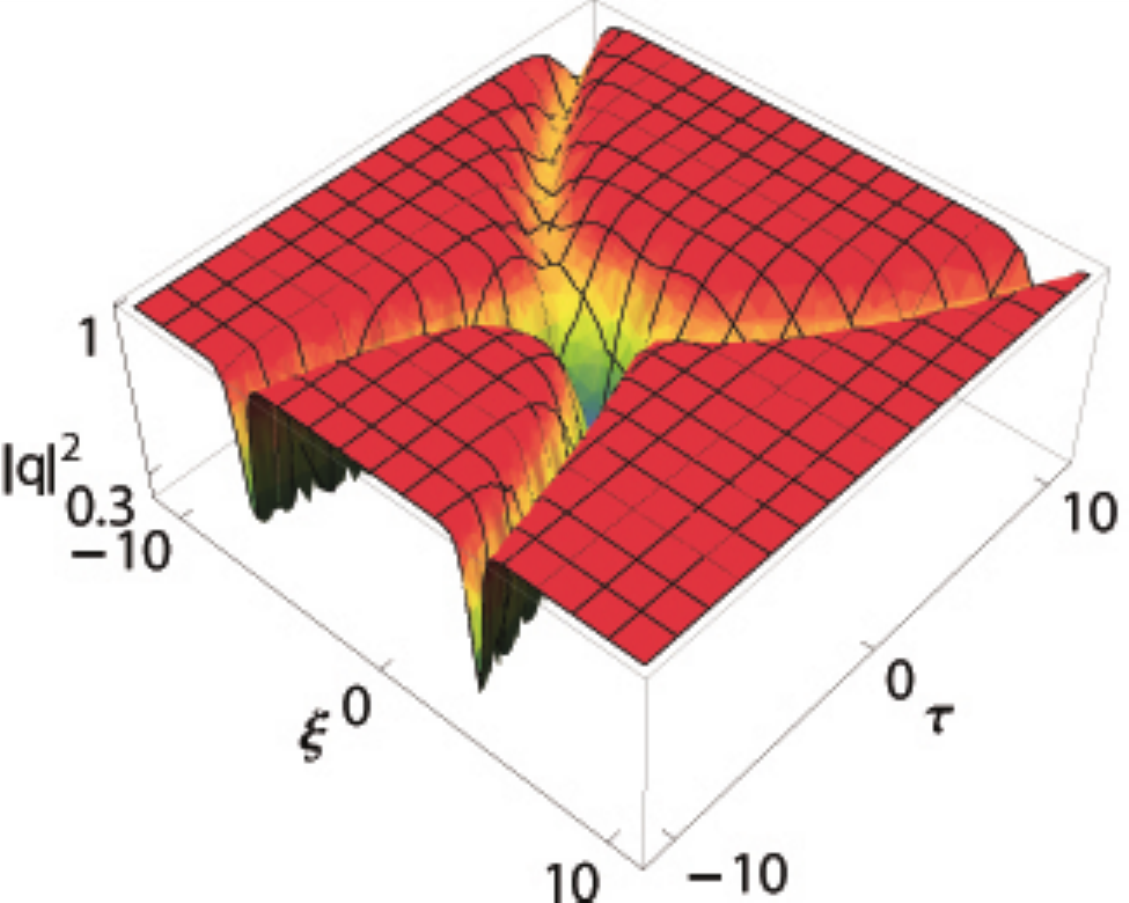}}
\caption{(Color online) The boomerang like dark solitons, (a) One-soliton, (b) Two-soliton. Other physical quantities are $k = a_0 = D(\xi) = 0.3+0.5\xi$, $p=0$,
$\delta=2$ and $\gamma=1$.}
\label{Boomerang}
\end{figure}

\begin{figure*}[ht!]
\subfloat[]{\includegraphics[width = 4cm]{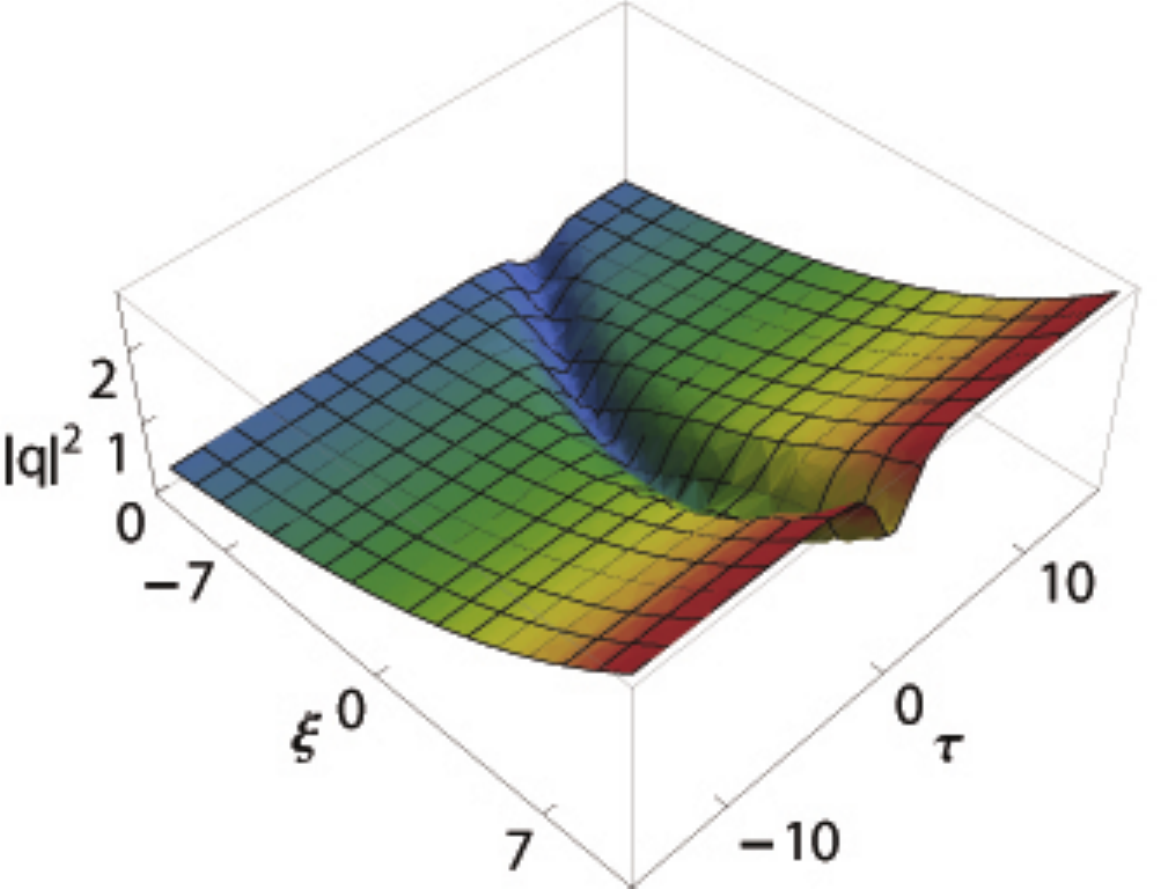}}
\subfloat[]{\includegraphics[width = 4cm]{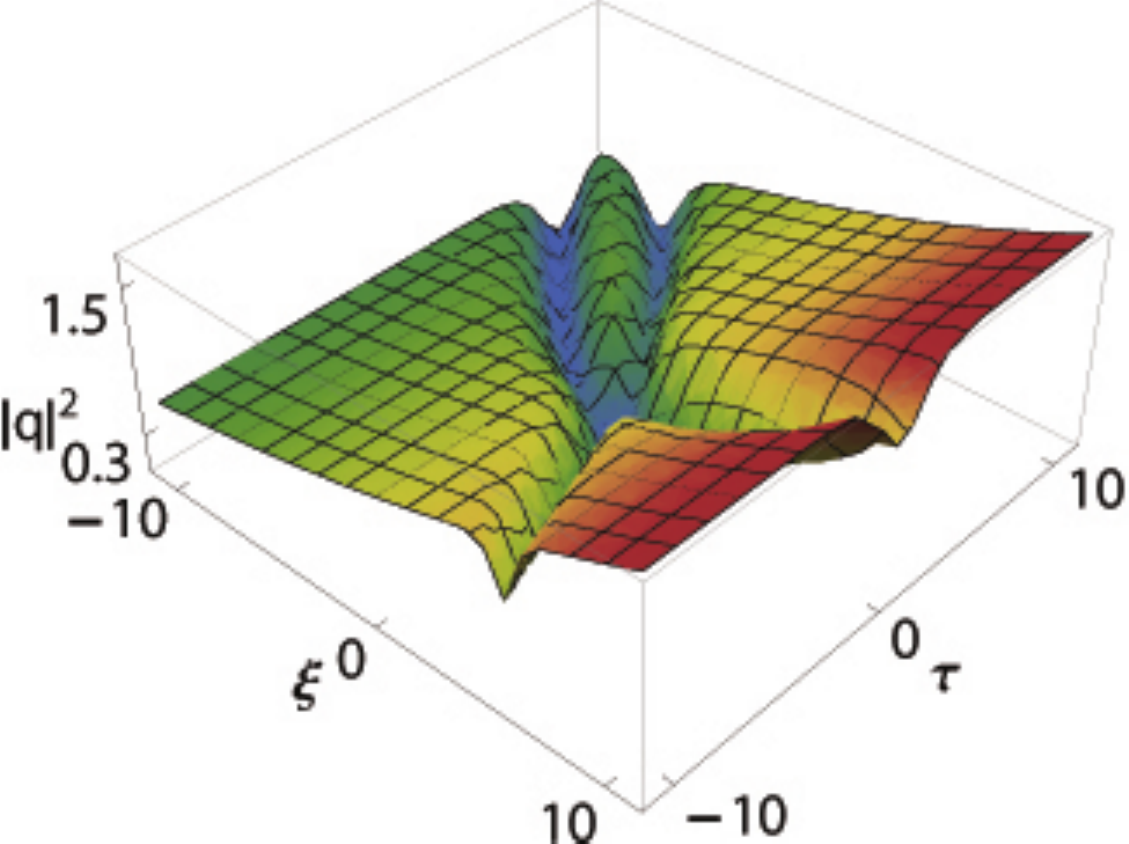}}
\subfloat[]{\includegraphics[width = 4cm]{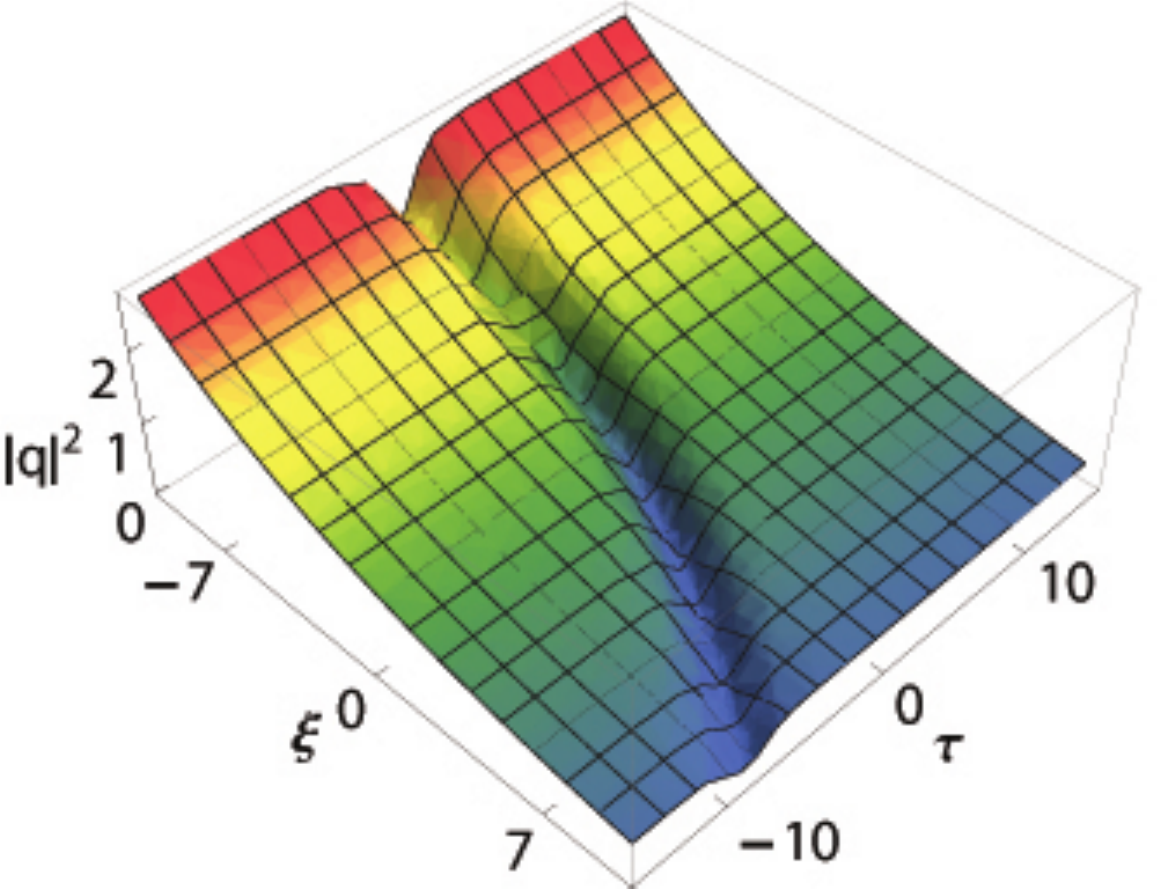}}
\subfloat[]{\includegraphics[width = 4cm]{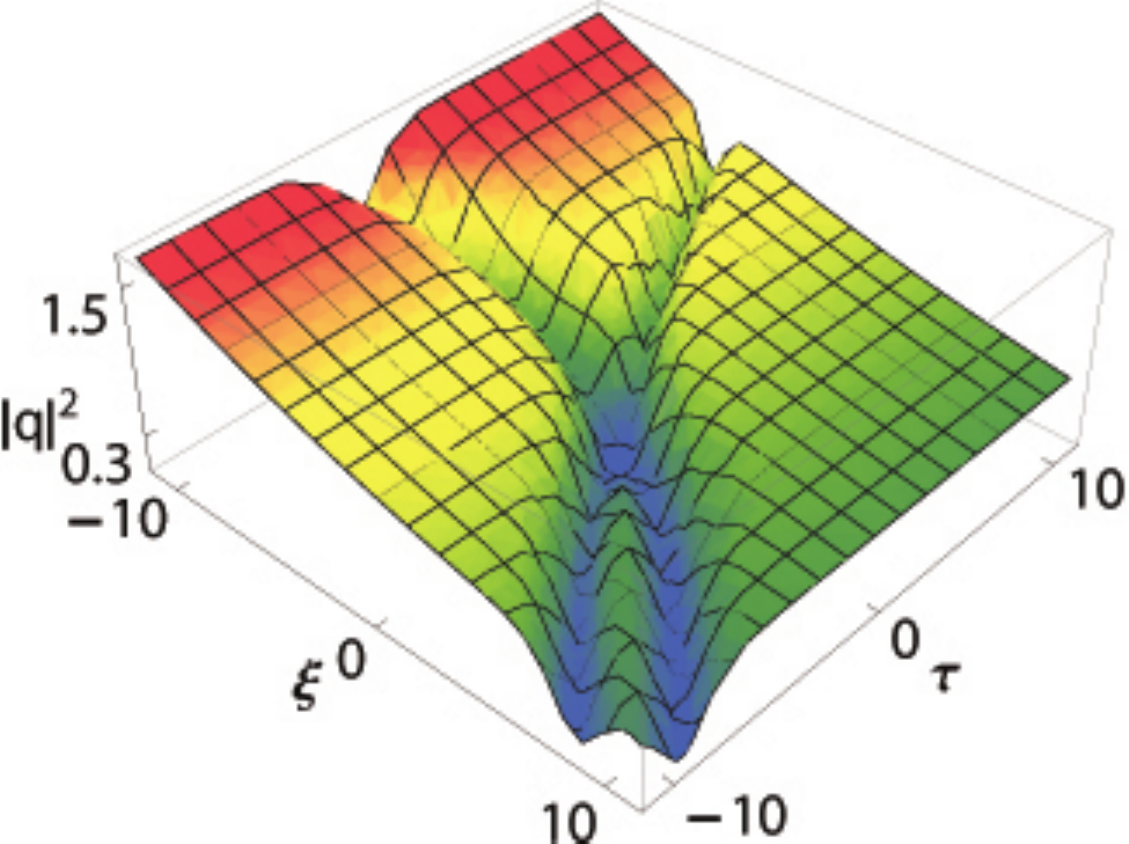}}
\caption{(Color online) The dark solitons propagation with gain, (a) One-soliton, (b) Two-soliton. The other parameters are $k = a_0 = D(\xi) = 1$,
$p=-0.05$,  $\delta= 2$ and $\gamma=0.2$. The dark solitons propagation with loss, (c) One-soliton (d) Two-soliton  with  $p=+0.05$.}
\label{gain}
\end{figure*}
\section{Results and discussions}
In the presented analytical work, we first investigated the constant propagation of dark soliton pulse in the homogeneous medium. In such system, the coefficient
corresponding to dispersion and nonlinearity remains constant. Using Hirota Bilinear method, the analytical dark soliton solution corresponding to one and two
are presented  graphically in Figs. \ref{dark one soliton} and \ref{dark two soliton}) via Eqs.  (\ref{one})and (\ref{two}) respectively. It is found that the
dark soliton propagates without deformation in such homogeneous system, and its amplitude and velocity remains constant.  In the following section, we examine
the dynamical evolution of dark soliton for different physical effects in inhomogenous fibers with variable coefficients, dispersion and nonlinearity.
\begin{figure*}[ht!]
\subfloat[]{\label{energy_gain}\includegraphics[width = 7cm]{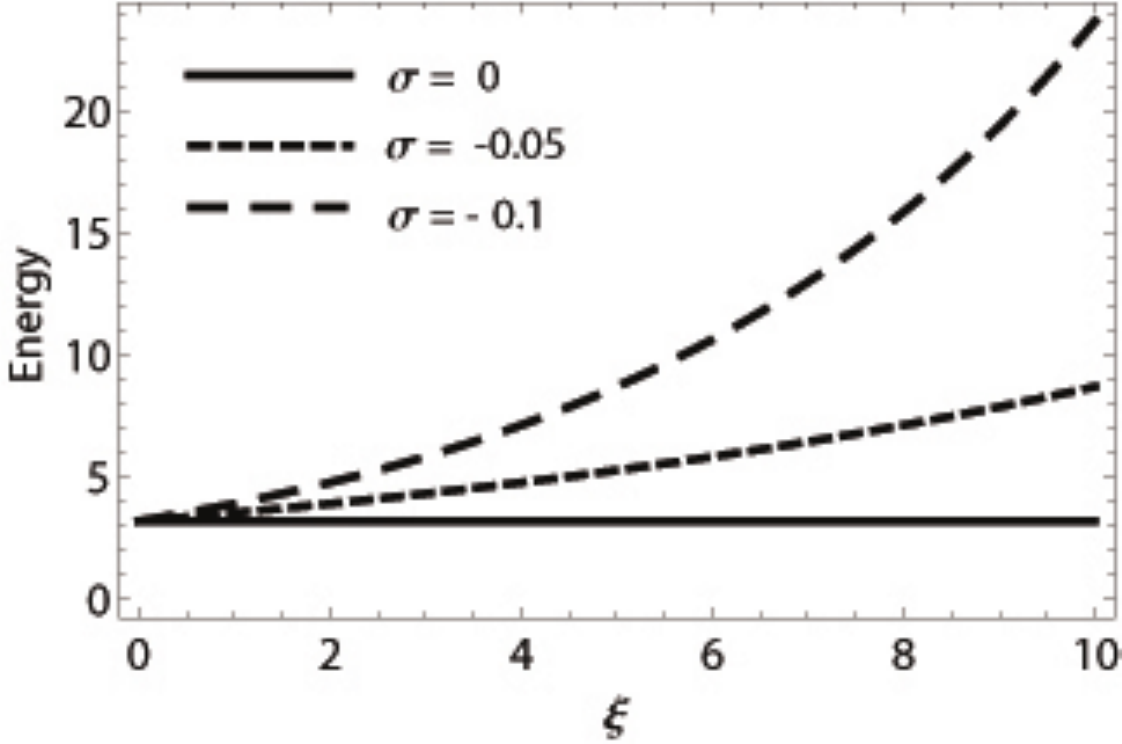}}\quad\quad\quad
\subfloat[]{\label{energy_loss}\includegraphics[width = 7cm]{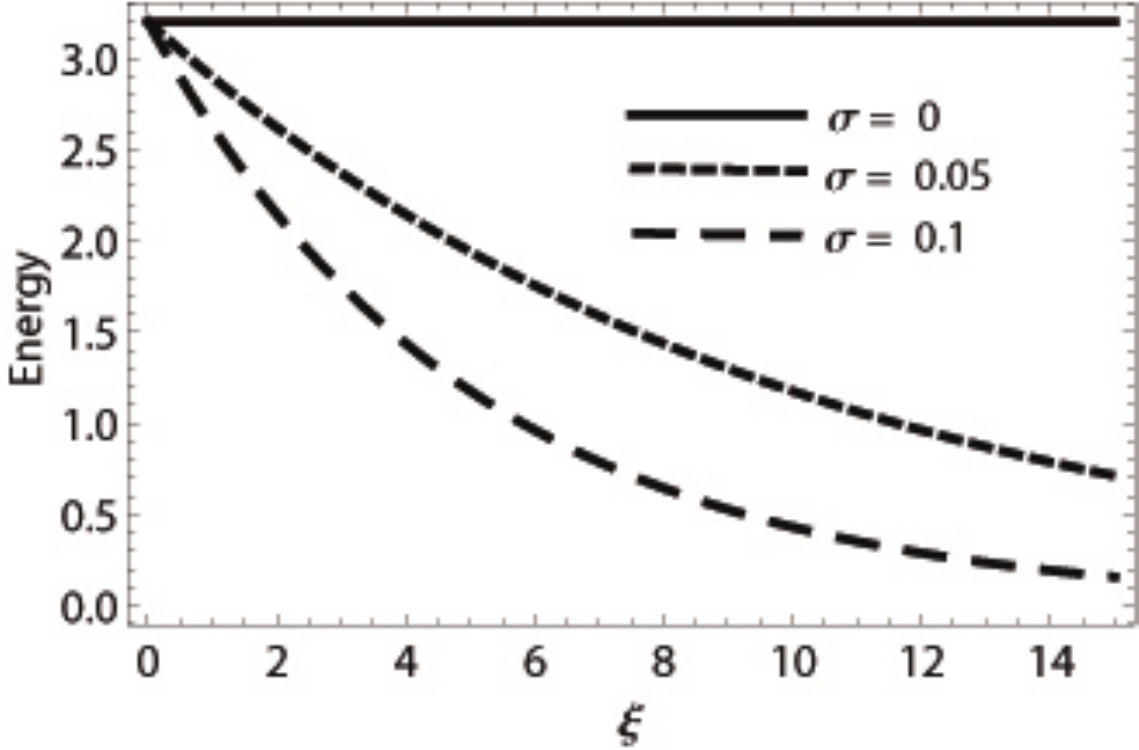}}
\caption{(Color online) Energy variation with gain/loss via the solution  (\ref{oneenergy}). (a) Gain with $p=-\sigma$, (b) Loss with $p=+\sigma$.  The other
relevant physical quantities are $ k = a_0 =\delta= D(\xi) = 1$,$k_1=1.5$ and $\gamma= 0.1$.}
\label{energy}
\end{figure*}

\begin{figure}[ht!]
\subfloat[]{\includegraphics[width = 4cm]{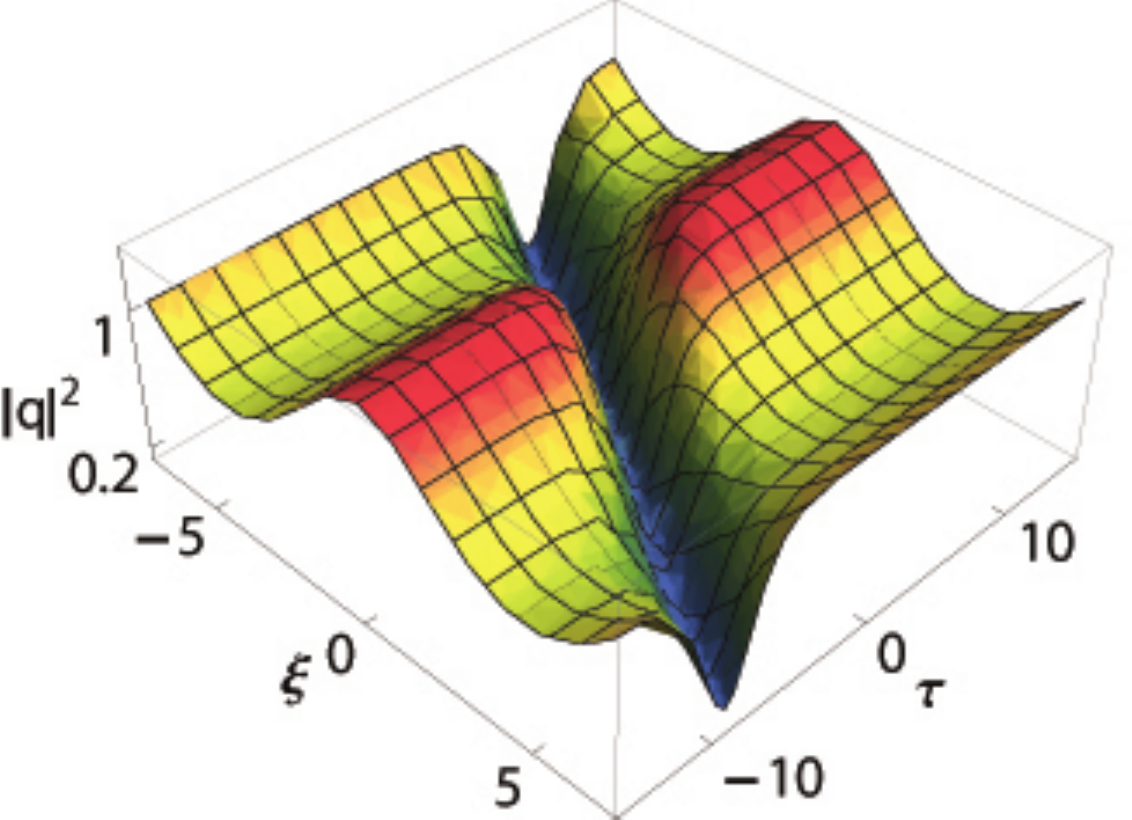}}
\subfloat[]{\includegraphics[width = 4cm]{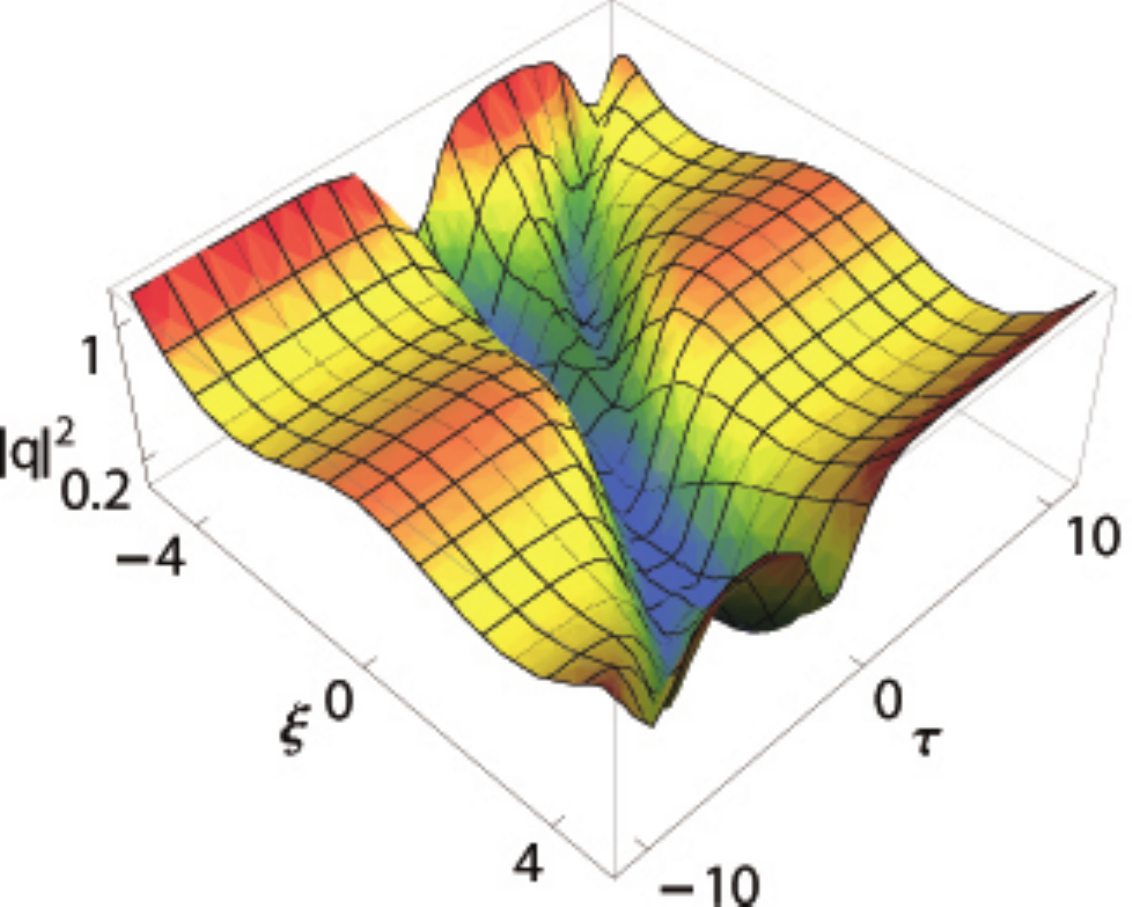}}
\caption{(Color online) The dark solitons propagation with periodic background, (a) One-soliton, (b) Two-soliton. The other physical quantities are $k = a_0 =
D(\xi) = 1$, $p=0.1 sin(0.7\xi)$,  $ R(\xi)=1$ and $ S(\xi)=0.5$.}
\label{oscillating}
\end{figure}

\subsection{Periodic varying dispersion and nonlinearity}
To study the dispersion-managed dark soliton by periodic perturbations, we consider a system with GVD parameter $D(\xi)$ and nonlinearity parameters $R(\xi)$ and
$S(\xi)$  as a trigonometric periodic function. In this case, the solitons are oscillating without any compression or broadening and the pulse peak position and
velocity vary periodically during the time of propagation. This kind of soliton is commonly called as a snaking soliton \cite{ref:61,ref:62,ref:63,ref:64}.
Similar type of inhomogeneous behavior is observed in two-soliton solution as well. In all the cases, the dispersion and nonlinearity parameters are
taken in the form of $a \cos (b\xi)$, where $a$  and $b$ are integers.  Fig. \ref{periodic}) represents the one and two  soliton pulse evolutions
with periodically varying effects. It shows that the amplitude, energy and pulse width remains constants during the propagation of the pulse down the fiber.

\begin{figure*}
\begin{center}
\subfloat[]{\label{dpowel}\includegraphics[width = 4cm]{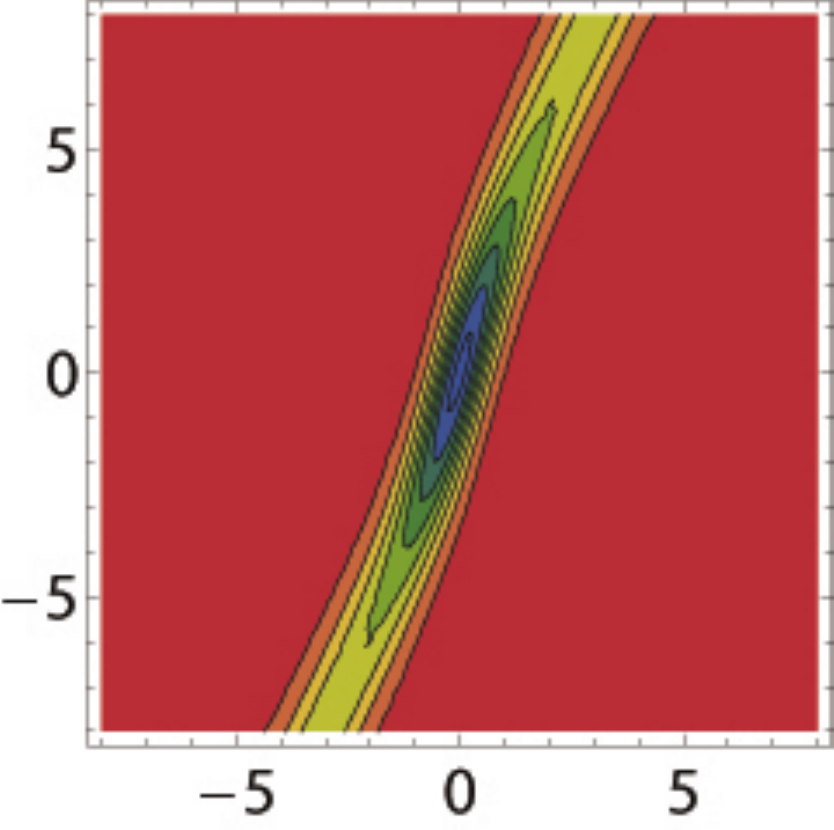}}
\subfloat[]{\label{dpobar}\includegraphics[width = 4cm]{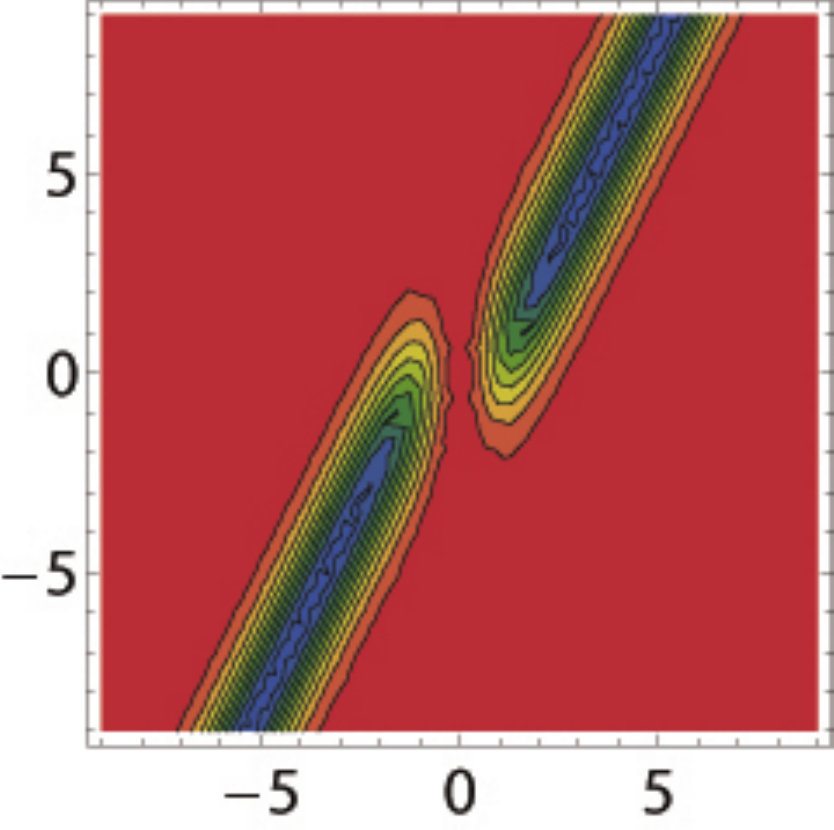}}
\subfloat[]{\label{nlobar}\includegraphics[width = 4cm]{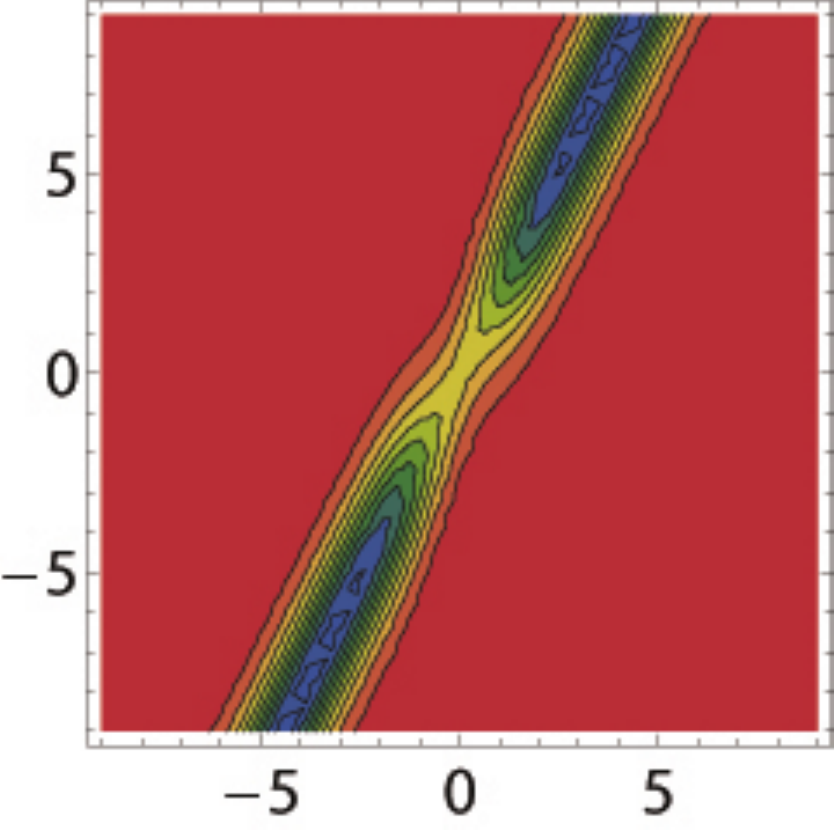}}
\subfloat[]{\label{nlowel}\includegraphics[width = 4cm]{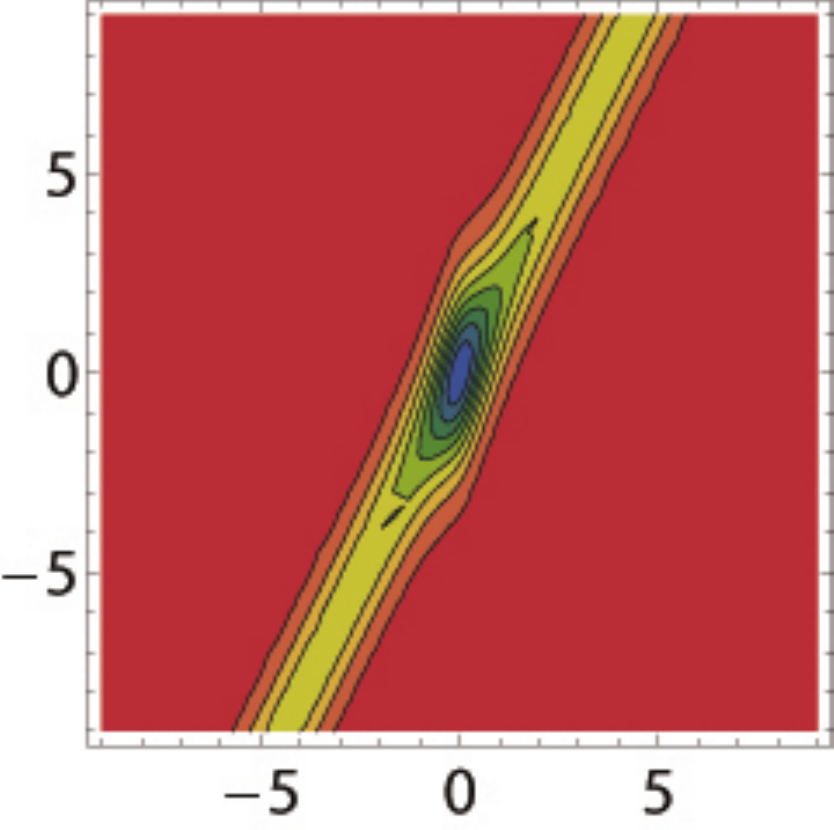}}
\caption{(Color online) Contour plot of nonlinear tunneling of dark one-soliton via solution (20). (a) Dispersion barrier with $D(\xi)= 1+ h sech[\xi-\xi_0]^2$,
$R=1$, $S=0.2$ and $h=0.5$.(b) Dispersion well with $h=-0.9$.(c) Nonlinear barrier with $D(\xi)= 1$,$R=1+ h sech[\xi-\xi_0]^2$,$S=0.2(1+ h sech[\xi-\xi_0]^2)$
and $h=1$.(d) Nonlinear well with $h=-0.5$. Other physical quantities are $k_1= a_0 =1$ and $ \xi_0=0$.}
\label{NL_one}
\end{center}
\end{figure*}

\subsection{Pulse Compression}
Pulse compression (PC) is an important technique to produce ultrashort pulse in nonlinear fiber. It a mechanism of shortening the duration of the pulse. Techniques like soliton effect, adiabatic pulse compression, self-similar methods are few of the most popular pulse compression techniques.
Generally, exponential dispersion and nonlinearity is preferred, as it found to compress soliton with a better a compression factor \cite{ref:65,ref:66}. In
similar lines with the earlier report, we consider the dispersion and nonlinearity parameters of the form $ c \exp(d\xi)$, where $c$ is the initial peak power
and $d$ is an integer. The  PC occurs, when the leading edge of the pulse is delayed by just the right amount to arrive nearly with the trailing edge
\cite{ref:4}. The PC of one and two  solitons are shown in the Fig. (\ref{Compression}).
\subsection{Boomerang Soliton}
To study the parabolic profile of dark soliton, we choose the dispersion and nonlinearity parameters as $e+f\xi$, where $e$ and $f$ are integers. During the
propagation, the soliton exhibits parabolic bending, and after the bending, the width of the soliton decreases gradually. Such type of soliton is commonly known as
Boomerang soliton \cite{ref:62,ref:63}. For the above choices the one and two-soliton solutions are shown in the Fig. (\ref{Boomerang})

\begin{figure*}
\begin{center}
\subfloat[]{\includegraphics[width = 4cm]{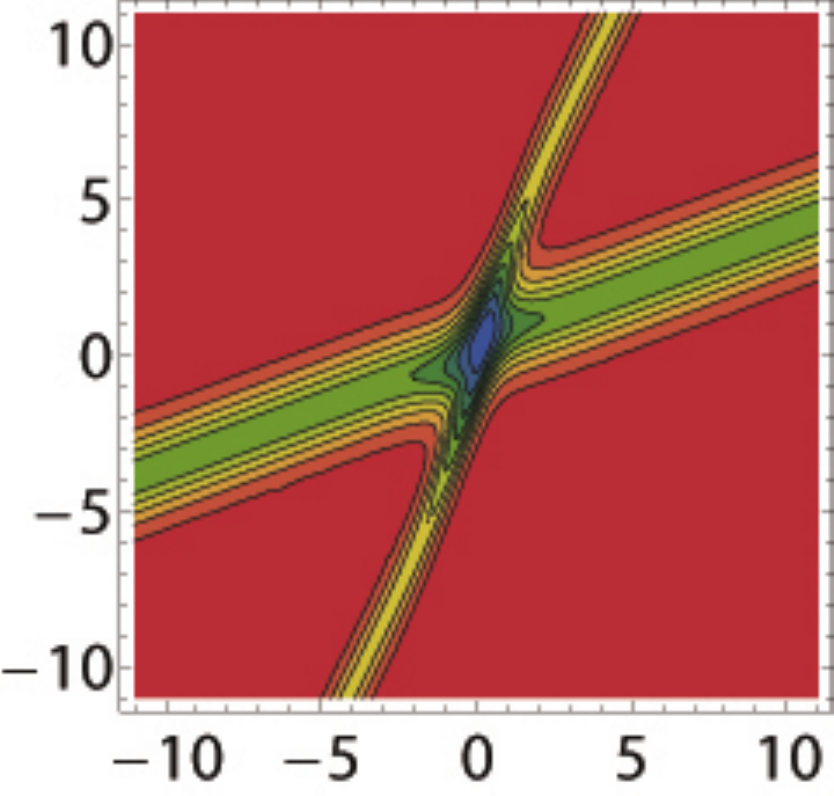}}
\subfloat[]{\includegraphics[width = 4cm]{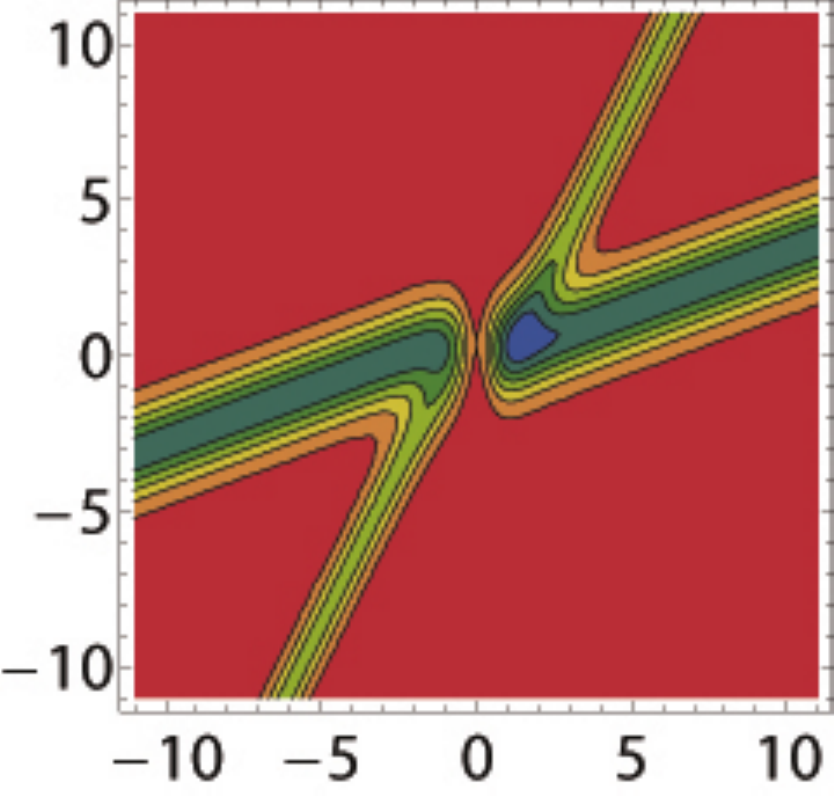}}
\subfloat[]{\includegraphics[width = 4cm]{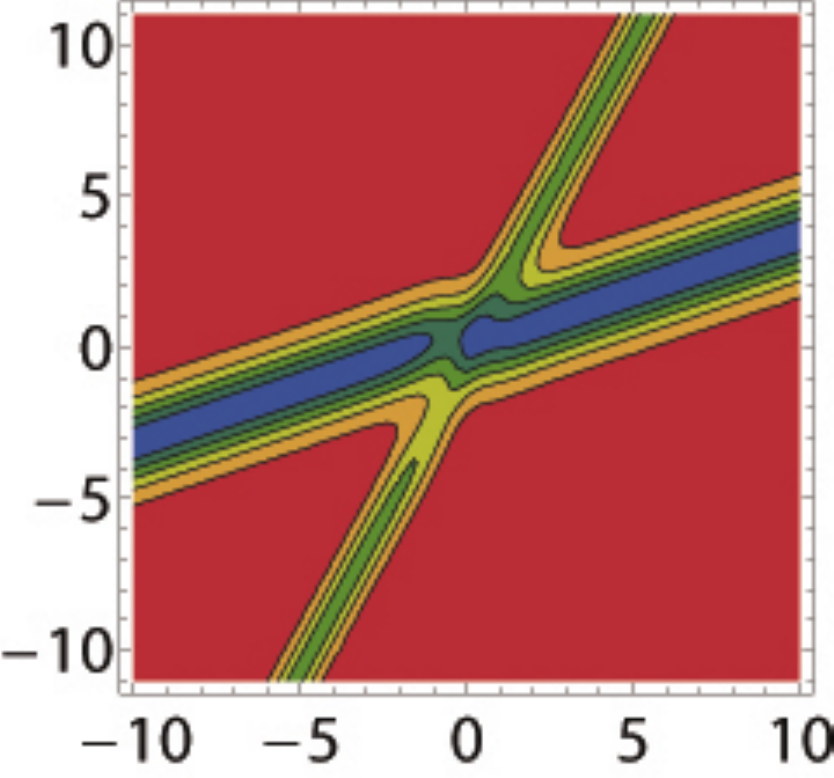}}
\subfloat[]{\includegraphics[width = 4cm]{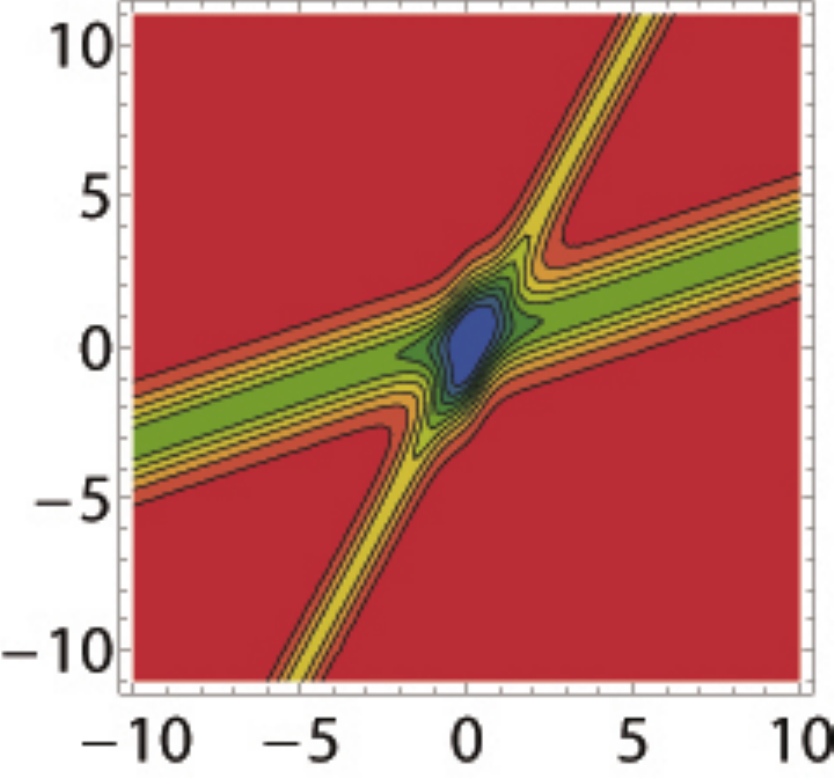}}
\caption{(Color online) Contour plot of nonlinear tunneling of dark two-soliton via solution (24). (a) Dispersion barrier with $D(\xi)= 1+ h sech[\xi-\xi_0]^2$,
$R=1$,$S=0.2$ and $h=0.7$.(b) Dispersion well with $h=-0.9$.(c) Nonlinear barrier with $D(\xi)= 1$,$R=1+ h sech[\xi-\xi_0]^2$,$S=0.2(1+ h sech[\xi-\xi_0]^2)$ and
$h=1$.(d) Nonlinear well with $h=-0.5$. Other physical quantities are $k_1=-0.8$, $k_2=0.8$, $a_0 =1$ and $ \xi_0=0$.}
\label{NL_two}
\end{center}
\end{figure*}

\subsection{Gain/Loss}
In the long distance optical fiber transmission system, the optical soliton will deform progressively as a result of fiber loss. In Eq. (\ref{vc-mnlse}) coefficient $p(\xi)$ plays
an important role in determining the amplification or absorption of the soliton pulse.  The coefficient $p(\xi)=0$ corresponds to the case of fibers without any
loss or gain.  When $p(\xi)$ is a constant value, say $\sigma$, the solution represents the propagation of soliton pulse in a medium with constant gain or loss.\\

Figs. \ref{gain}  illustrate the propagation of one and two solitons in a medium with gain/loss. When $\sigma<0(\sigma>0)$, the pulse
undergoes the amplification (compression), and their amplitude of the pulse increases (decreases) as it propagates down the fiber. By varying the
coefficient $p(\xi)$, the soliton amplification or absorption can be controlled. Fig. \ref{energy}, represents the variation of dark soliton energy for some representative values of gain (Fig. \ref{energy_gain}) and loss (Fig. \ref{energy_loss}).  It is quite evident that the energy monotonously increases (decreases) with  gain (loss).

The gain/loss coefficient $p(\xi)$ can significantly influences the shape of the pulse background, in many recent works,  different type of background profiles were
considered \cite{ref:67}. Here, we demonstrated that the background undergoes a periodic oscillation. Fig. \ref{oscillating} shows the evolution of one- and two- solitons in a periodic background with $p(\xi) = g \sin(h\xi)$, where $g$ and $h$ are integers.

\begin{figure*}
\subfloat[]{\label{dpoweld}\includegraphics[width = 4cm]{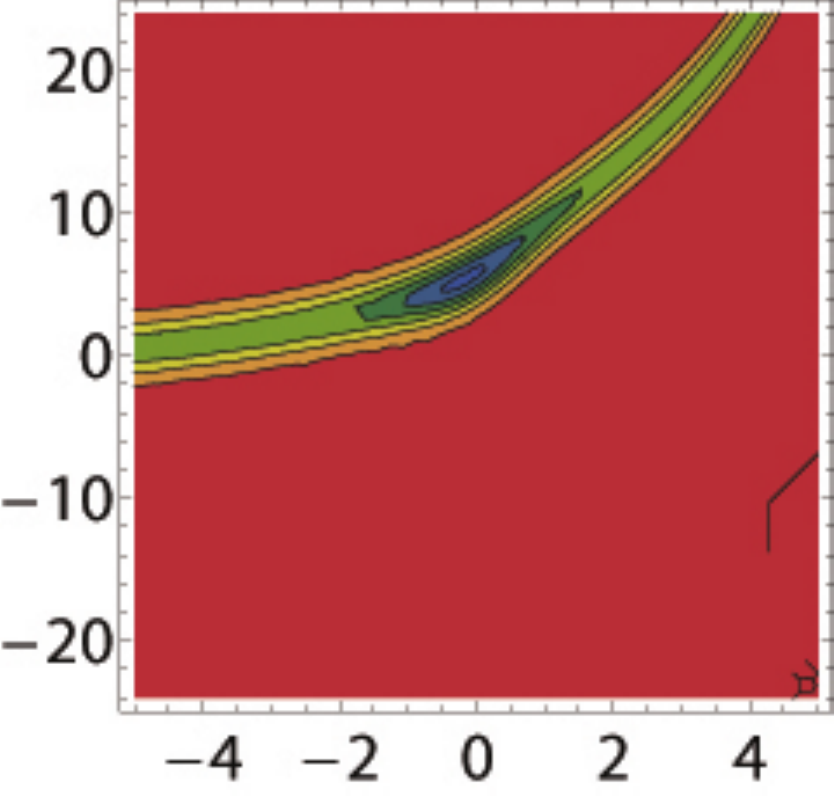}}
\subfloat[]{\label{dptweld}\includegraphics[width = 4cm]{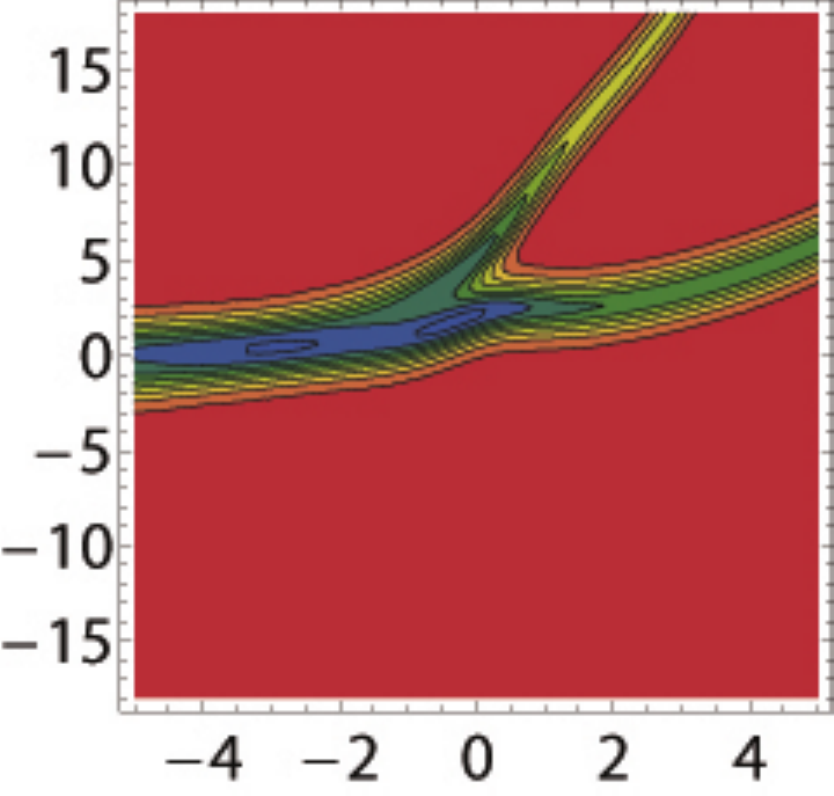}}
\subfloat[]{\label{dpobard}\includegraphics[width = 4cm]{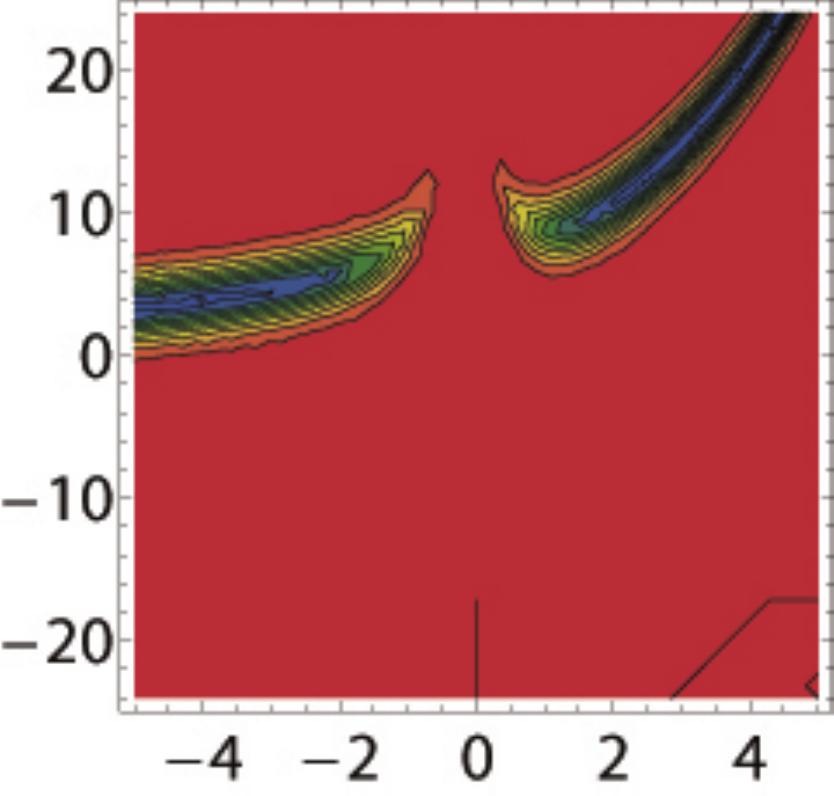}}
\subfloat[]{\label{dptbard}\includegraphics[width = 4cm]{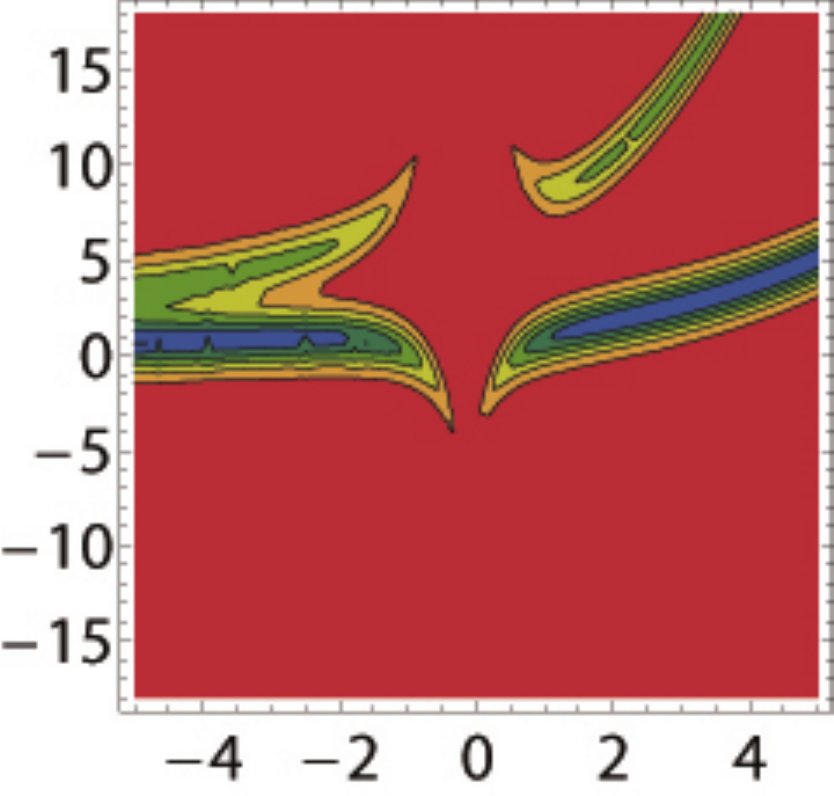}}
 \caption{(Color online) Contour plot of nonlinear tunneling with exponential background.  Dispersion barrier of (a) One soliton, (b) Two soliton,   with
 $D(\xi)=  d_0 e^{-r\xi} + h sech[\xi-\xi_0]^2$,$R=R_0 e^{-r\xi}$,$S=S_0 e^{-r\xi}$ and $h=0.7$. Dispersion well of (c) One soliton, (d) Two soliton,   with the
 same as for (a-b) except that  $h=-0.9$. Other physical quantities are $d_0 = R_0=1$  $S_0=0.2$,$r=-0.3$  $k_1=-0.8$, $k_2=0.8$, $k_3=-0.7$ and $ \xi_0=0$.}
\label{NLexponential}
\end{figure*}
\section{Nonlinear tunneling effect}
In the previous sections, we discussed about the impact of various physical effects and inhomogenous parameters in the one and two solitons. Now, we
intent to investigate one of the dramatic nonlinear effects, known as the nonlinear tunneling (NL). Recently, many leading research works have been devoted to
investigate the tunneling of solitons in different physical systems \cite{ref:48,ref:49,ref:50,ref:51,ref:52,ref:53,ref:54,ref:55,ref:56,ref:57,ref:58,ref:59}.
All pioneering works have shown that the soliton can pass through the barrier without loss under a special conditions, which depends on the ratio between the
height of the barrier and the amplitude of the soliton.  The NL tunneling of soliton may create a new field of interest and feature wide applications in
all-optical switches and logic circuits.
\subsubsection{Nonlinear tunneling without exponential background}
To investigate the  NL of Vc-MNLS dark soliton propagating through the dispersion barrier or well, we choose the dispersion and nonlinear parameter as follows:

\begin{align*}
D(\xi)&=r_0\pm h\, sech^2(c(\xi-\xi_0)) \\
R(\xi)&=R_0   \\
S(\xi)&=S_0
\end{align*}
 In the above expressions, $h$ and $c$ represent the height and width of the barrier. $\xi_0$ represents the longitudinal co-ordinate indicating the location of
 the dispersion barrier/well.  $D_0$, $R_0 $ and $S_0 $ are constant parameters. Here the positive or the negative sign of $\pm h$ denotes the barrier or the
 well. If $h=0$, the soliton is said to propagate through a homogenous fiber.\\

To investigate the soliton propagation through the nonlinear barrier or well, we consider the variable coefficients as follows:
  \begin{align*}
  D(\xi)&=D_0  \\
   R(\xi)&= R_0(r_0\pm h\, sech^2(c(\xi-\xi_0))\\
   S(\xi)&= S_0(r_0\pm h\, sech^2(c(\xi-\xi_0))
   \end{align*}

When the dark soliton is passing through the dispersion barrier, the intensity of the soliton grows and forms a peak at $ \xi=\xi_0$.  After passing through the
barrier, the pulse retains its original shape as illustrated in  Fig. \ref{dpowel}. For the case of dispersion well, the amplitude of the soliton vanishes and a
valley is formed at $\xi=\xi_0$; after the tunneling, solitons are restored to their original shape as shown in Fig. \ref{dpobar}. On the other hand for
nonlinear barrier/well, a valley is formed for nonlinear barrier and a peak is formed for well as evident from the Figs. \ref{nlobar} and \ref{nlowel},
respectively. In similar lines with one-soliton case, the NL tunneling of two solitons is demonstrated in the Figs. \ref{NL_two}.

Due to the existing region of dark soliton ($ 0\leq A < a_0$), we can found that the height of the barrier (h) and amplitude of pulse (A) have a mutual relation  in given soliton solutions. Thus we have studied the tunneling effect with suitable parametric choice of $h$. To investigate the dark soliton propagation through the
dispersion barrier or well, we obtained a condition, where $h > 0$ indicates the dispersion barrier, and $-1 < h < 0$ represents the dispersion well.  Similarly,
in the case of nonlinear barrier or well, we also obtained a condition, where $h > 0$ indicates the nonlinear barrier, and $-1 < h < 0$ represents the nonlinear
well.
\subsection{Nonlinear tunneling with exponential background}
Now, we consider the tunneling effect with exponential background. This case is of particular importance, because, pulse tunneling through the exponential
background generally results in the compression of the pulse. To investigate this special case, we consider the dispersion and nonlinear parameter as follows:
    \begin{align*}
    D(\xi)&=D_0 exp(-r\xi)\pm h\, sech^2(c(\xi-\xi_0)) \\
    R(\xi)&=R_0 exp(-r\xi) \\
    S(\xi)&=S_0 exp(-r\xi)
    \end{align*}
Here, $r$ in the above expression represents the decaying parameter, which accounts for the exponential decay.  Figs. (\ref{dpoweld} - \ref{dptweld}), represents
the dark one and two  solitons tunneling through dispersion barrier with exponential decay. It is observed that the amplitude of the soliton increases at
$\xi=\xi_0$, and after tunneling through the barriers, the width of the soliton decreases gradually during propagation. Similarly, Figs. (\ref{dpobard} -
\ref{dptbard}) illustrates the  dark soliton tunneling through dispersion well with exponential background for the case of one and two   solitons.  As it is
evident that the amplitude of the soliton vanishes at $\xi=\xi_0$ and after emerging from the well, the soliton width compresses. From this result, one can infer
that the input pulse can be compressed to a desired extent in a controllable manner by the proper choice of barrier or well parameters.
\section{Summary and conclusion}
In this paper, we have investigated a Vc-MNLS model with distributed dispersion, SPM, SS and linear gain/loss, which describes the dynamics of ultrashort pulse
propagation in the inhomogeneous fiber systems. Using Hirota's bilinear method, we analytically derived the exact one and two dark soliton solutions. We illustrated the effect of self-steepening such as shock wave formation during the propagation, which has been confirmed through direct numerical simulation. In order to validate the stability of the soliton solution, we numerically perform the stability analysis in the presence of photon noise, and our simulation illustrates the stable propagation of the dark soliton pulse. The collision behaviors of the dark soliton pulses in inhomogeneous fibers have also been discussed through asymptotic analysis.

For better insight about the effect of inhomogeneity, we exclusively studied the dynamical behavior of dark soliton with different physical effects up to the level of two-dark soliton interactions.  In particular, we focused on the nonlinear tunneling of dark soliton through barrier/well.  It has been found that the intensity of the tunneling soliton either forms a peak or valley and retains its shape after tunneling through barrier/well. We also identified the tunneling of dark soliton with exponential background tends to compress the pulse. Thus, in this paper, we attempt to give a complete study about the dark soliton dynamics in the Vc-MNLS model, by incorporating most of the physical effects. We believe the aforementioned results of the paper can serve as a reference for many future studies related to dark soliton.
\section{Acknowledgements}
KP thanks DST, CSIR, NBHM, IFCPAR and DST-FCT Government of India, for the financial support through major projects. N. M. Musammil thanks UGC – MANF, Government
of India, for the financial support through Junior Research fellow. K.Nithyanandan thanks CNRS for post doctoral fellowship at Universite de Bourgogne, Dijon, France.
\bibliographystyle{99}

\end{document}